\documentstyle[12pt,epsf,a4]{article}
\newcommand{\bv}{\begin{verbatim}}

\newcommand{\ev}{\end{verbatim}}

\newcommand{\smM}{\mbox{\small{\it M}}}

\renewcommand{\thefootnote}{\fnsymbol{footnote}}
\newcommand{\li}{\mathop{{\mbox{Li}}_4}\nolimits}

\newcommand{\gsim}{\;\rlap{\lower 3.5 pt \hbox{$\mathchar \sim$}} \raise 1pt
 \hbox {$>$}\;}
\newcommand{\lsim}{\;\rlap{\lower 3.5 pt \hbox{$\mathchar \sim$}} \raise 1pt
 \hbox {$<$}\;}

\newcommand{\lmM}{l_{\mu M}}

\begin{document}    

\title{\vskip-3cm{\baselineskip14pt
\centerline{\normalsize\hfill DESY 00--034}
\centerline{\normalsize\hfill TTP00--05}
\centerline{\normalsize\hfill hep-ph/0004189}
\centerline{\normalsize\hfill March 2000}
}
\vskip.7cm
{\tt RunDec}: a {\tt Mathematica} package for running and decoupling of the
strong coupling and quark masses
\vskip.3cm
}
\author{
{K.G. Chetyrkin}$^{a,}$\thanks{Permanent address:
Institute for Nuclear Research, Russian Academy of Sciences,
60th October Anniversary Prospect 7a, Moscow 117312, Russia.}
\,,
{J.H. K\"uhn}$^a$
\,and
{M. Steinhauser}$^b$
  \\[3em]
  { (a) Institut f\"ur Theoretische Teilchenphysik,}\\
  { Universit\"at Karlsruhe, D-76128 Karlsruhe, Germany}
  \\[.5em]
  { (b) II. Institut f\"ur Theoretische Physik,}\\ 
  { Universit\"at Hamburg, D-22761 Hamburg, Germany}
}
\date{}
\maketitle

\begin{abstract}
\noindent
In this paper the formulae are collected which are needed
for the computation of the strong coupling constant and
quark masses at different energy scales and for different number of active
flavours. All equations contain the state-of-the-art QCD corrections
up to three- and sometimes even four-loop order.
For the practical implementation {\tt Mathematica} is used and a package
containing useful procedures is provided. 

\vspace{.2cm}

\noindent
PACS numbers: 12.38.Aw 12.38.-t 14.65.-q

\end{abstract}

\thispagestyle{empty}
\newpage
\setcounter{page}{1}

\renewcommand{\thefootnote}{\arabic{footnote}}
\setcounter{footnote}{0}


\section*{Program Summary}

\begin{itemize}

\item[]{\it Title of program:}
  {\tt RunDec}

\item[]{\it Available from:}\\
  {\tt
  http://www-ttp.physik.uni-karlsruhe.de/Progdata/ttp00/ttp00-05/
  }

\item[]{\it Computer for which the program is designed and others on which it
    is operable:}
  Any work-station or PC where {\tt Mathematica} is running.

\item[]{\it Operating system or monitor under which the program has been
    tested:} 
  UNIX, Mathematica~4.0

\item[]{\it No. of bytes in distributed program including test data etc.:}
  $65000$

\item[]{\it Distribution format:} 
  ASCII

\item[]{\it Keywords:} 
  Quantum Chromodynamics, running coupling constant,
  running quark mass, on-shell mass, $\overline{\rm MS}$ mass,
  decoupling of heavy particles

\item[]{\it Nature of physical problem:}
  The values for the coupling constant of Quantum Chromodynamics,
  $\alpha_s^{(n_f)}(\mu)$, actually
  depends on the considered energy scale, $\mu$, and the number of active
  quark flavours, $n_f$. The same applies to light quark masses,
  $m_q^{(n_f)}(\mu)$,
  if they are, e.g., evaluated in the $\overline{\rm MS}$ scheme.
  In the program {\tt RunDec} all 
  relevant formulae are collected and various procedures are provided
  which allow for a convenient evaluation of $\alpha_s^{(n_f)}(\mu)$
  and $m_q^{(n_f)}(\mu)$ using the state-of-the-art correction terms.

\item[]{\it Method of solution:}
  {\tt RunDec} uses {\tt Mathematica} functions to perform the
  different mathematical operations.
  
\item[]{\it Restrictions on the complexity of the problem:}
  It could be that for an unphysical choice of the input parameters the 
  results are nonsensical.

\item[]{\it Typical running time:}
  For all operations the running time does not exceed a few seconds.

\end{itemize}

\section{Introduction}

Quantum Chromodynamics (QCD) is nowadays well established as the  theory of
strong interaction within the Standard Model of elementary particle physics.
In  recent years there has been  a wealth of theoretical results  
(for a review see~\cite{ckk96}).
At the same time  perturbative QCD has been extremely successful in
describing  the experimental data with  high precision. 

The fundamental quantity of QCD is the so-called beta function which
connects the value of the strong coupling constant, $\alpha_s(\mu)$,
at different energy scales $\mu$.  It is thus particularly important
to know the beta function as precise as
possible. In~\cite{RitVerLar97} the four-loop corrections were
evaluated allowing for a consistent running at order $\alpha_s^4$.  In
the majority of all computations performed in QCD the $\overline{\rm
MS}$ renormalization scheme~\cite{BarBurDukMut78} is adopted. In this
scheme the Appelquist-Carazzone decoupling theorem~\cite{AppCar75} is
not directly applicable. When crossing flavour
thresholds, it is thus important to perform the decoupling ``by
hand''.  In order to be consistent,  four-loop running must go along
with the three-loop decoupling relation which was evaluated
in~\cite{CheKniSte97}.

Similar considerations are also valid for   quark masses.
Also here the renormalization group function is available up to the
four-loop level~\cite{Che97LarRitVer972} and the corresponding
decoupling relation up to order $\alpha_s^3$~\cite{CheKniSte98}
(see also~\cite{Rod9397}).

In this paper all relevant formulae are collected which are necessary for the
running and decoupling of $\alpha_s$ and  for quark masses.
Their proper use is discussed and easy-to-use {\tt Mathematica}~\cite{math}
procedures
collected in the package {\tt RunDec}
are provided. Their handling is described and examples are given.

The outline of the paper is as follows. In the next Section the 
formulae are presented which are
needed for the running of the strong coupling constant up to the
four-loop level. The corresponding  equations for the quark masses are 
presented in Section~\ref{sec:masses}. In addition the conversion formulae
between the $\overline{\rm MS}$ and on-shell scheme are discussed in some
detail. Section~\ref{sec:dec} is concerned with the decoupling of the strong
coupling and  quark masses.
Finally, in Section~\ref{sec:desc}, the most important procedures of the
package {\tt RunDec} are described in an easy-to-use way.
For most practical applications they should be sufficient.
In the Appendix the complete  collection of procedures is given.


\section{Running strong coupling constant}

The beta function governing the running of the coupling constant
of QCD is defined through
\begin{eqnarray}
  \mu^2\frac{{\rm d}}{{\rm d}\mu^2}
  \frac{\alpha_s^{(n_f)}(\mu)}{\pi}
  &=&
  \beta^{(n_f)}\left(\alpha_s^{(n_f)}\right)
  \,\,=\,\,
  - \sum_{i\ge0}
  \beta_i^{(n_f)}\left(\frac{\alpha_s^{(n_f)}(\mu)}{\pi}\right)^{i+2} 
  \,,
  \label{eq:defbeta}
\end{eqnarray}
where $n_f$ is the number of active flavours.
The coefficients are given by~\cite{gro,jon,tar,RitVerLar97}
\begin{eqnarray}
  \beta_0^{(n_f)} &=&\frac{1}{4}\left[ 11 - \frac{2}{3} n_f\right]
  \,,
  \nonumber\\
  \beta_1^{(n_f)} &=&\frac{1}{16}\left[ 102 - \frac{38}{3} n_f\right]
  \,,
  \nonumber \\
  \beta_2^{(n_f)} &=&\frac{1}{64}\left[\frac{2857}{2} - \frac{5033}{18} n_f
    + \frac{325}{54} n_f^2\right]
  \,,
  \nonumber \\
  \beta_3^{(n_f)} &=&\frac{1}{256}\left[  \frac{149753}{6} + 3564 \zeta_3 
    + \left(- \frac{1078361}{162} - \frac{6508}{27} \zeta_3 \right) n_f
  \right.\nonumber\\&&\mbox{}
  + \left( \frac{50065}{162} + \frac{6472}{81} \zeta_3 \right) n_f^2
  +\left.  \frac{1093}{729}  n_f^3\right]
  \,.
  \label{eq:betafct}
\end{eqnarray}
$\zeta$ is Riemann's zeta function, with values $\zeta_2=\pi^2/6$
and $\zeta_3\approx1.202\,057$.
It is convenient to introduce the following notation:
\begin{eqnarray}
  b_i^{(n_f)} &=& \frac{\beta_i^{(n_f)}}{\beta_0^{(n_f)}}
  \,,
  \nonumber\\
  a^{(n_f)}(\mu) &=& \frac{\alpha^{(n_f)}(\mu)}{\pi}
  \,.
  \label{eq:bidef}
\end{eqnarray}
In the following
the labels $\mu$ and $n_f$ are omitted if confusion is impossible.

Integrating Eq.~(\ref{eq:defbeta}) leads to
\begin{eqnarray}
  \ln\frac{\mu^2}{\Lambda^2}&=&\int\frac{da}{\beta(a)}
\nonumber
\\
  &=&\frac{1}{\beta_0}\left[\frac{1}{a}+b_1\ln a+(b_2-b_1^2)a
    + \left(\frac{b_3}{2}-b_1b_2+\frac{b_1^3}{2}\right)a^2\right]+C,
  \label{eq:lamexpl}
\end{eqnarray}
where an expansion in $a$ has been performed.
The integration constant is conveniently split into $\Lambda$, the
so-called asymptotic scale parameter, and $C$.
The conventional $\overline{\rm MS}$ definition of $\Lambda$, which we shall
adopt in the following, corresponds to choosing
$C=(b_1/\beta_0)\ln\beta_0$~\cite{BarBurDukMut78,FurPet82}.

Iteratively solving Eq.~(\ref{eq:lamexpl}) yields~\cite{CheKniSte97}
\begin{eqnarray}
  a&=&\frac{1}{\beta_0L}-\frac{b_1\ln L}{(\beta_0L)^2}
  +\frac{1}{(\beta_0L)^3}\left[b_1^2(\ln^2L-\ln L-1)+b_2\right]
  \nonumber\\
  &&\mbox{}+\frac{1}{(\beta_0L)^4}\left[
    b_1^3\left(-\ln^3L+\frac{5}{2}\ln^2L+2\ln L-\frac{1}{2}\right)
    -3b_1b_2\ln L+\frac{b_3}{2}\right],
  \label{eq:alp}
\end{eqnarray}
where $L=\ln(\mu^2/\Lambda^2)$ and terms of ${\cal O}(1/L^5)$ have been
neglected.
$\Lambda$ is defined in such a way that
Eq.~(\ref{eq:alp}) does not contain a term proportional to
$({\rm const.}/L^2)$~\cite{BarBurDukMut78}.

The canonical way to compute $a(\mu_2)$ when $a(\mu_1)$ is given for a fixed
number of flavours is as follows: 
\begin{enumerate}
\item
  Determine $\Lambda$. There are several possibilities to do this.
  One could, e.g., use the explicit solution given in
  Eq.~(\ref{eq:lamexpl}). Another possibility is the use of (\ref{eq:alp}) and
  solve the equation iteratively for $\Lambda$.
  Furthermore the first line of~(\ref{eq:lamexpl}) could be used and
  the integral could be solved numerically 
  without performing any expansion in $\alpha_s$. We will see in the examples
  below that the numerical differences are small.
\item
  $a(\mu_2)$ is computed with the help of Eq.~(\ref{eq:alp}) where the value
  of $\Lambda$ is inserted and $\mu$ is set to $\mu_2$.
\end{enumerate}

It is also possible to avoid the introduction of $\Lambda$ in intermediate
steps and to solve the differential equation~(\ref{eq:defbeta})
numerically using  $a(\mu)|_{\mu=\mu_1}=a(\mu_1)$ as initial condition.
This convention  requires the knowledge of both $\alpha_s$ and the
scale $\mu$ in order to determine $\alpha_s$ at the new scale.
Frequently, $\mu = M_Z$ is  used as reference scale.

On the other hand
$\Lambda$ plays the role of an universal parameter which at the same time sets
the characteristic scale of QCD.

At this point it is instructive to consider an example.  Let us assume
that $\alpha_s$ is given at the $Z$-boson scale:
$\alpha_s^{(5)}(M_Z)=0.118$.  Let us further assume that it is
determined from the experiment with three-loop accuracy, which means
that in the $\beta$ function~(\ref{eq:betafct}) only the coefficients
up to $\beta_2$ are considered and $\beta_3$ is neglected.  Let us now
evaluate the strong coupling at the scale $\mu=M_b$ and compare the
results obtained with the different strategies outlined above. In the
following a
possible {\tt Mathematica} session is shown. {\tt NumDef} is a set of {\tt
  Mathematica} rules which assigns typical values to the physical parameters
used in our procedures. The numbers used in this paper can be found in
Eq.~(\ref{eq:numdef}) and the procedures are described in the Appendix. 
\begin{verbatim}
In[1]:= <<RunDec.m;
\end{verbatim}
Comment: evaluation of $\Lambda$ from $\alpha_s^{(5)}(M_Z)$ based on 
the explicit solution, Eq.~(\ref{eq:lamexpl}), and subsequent  
evaluation of $\alpha_s^{(5)}(M_b)$ from $\Lambda$ based on Eq.~(\ref{eq:alp}).
\begin{verbatim}
In[2]:= lamex = LamExpl[asMz/.NumDef,Mz/.NumDef,5,3]

Out[2]= 0.208905

In[3]:= AlphasLam[lamex,Mb/.NumDef,5,3]

Out[3]= 0.216610
\end{verbatim}
Comment: evaluation of $\Lambda$ from $\alpha_s^{(5)}(M_Z)$ based on 
Eq.~(\ref{eq:alp}), and subsequent  
evaluation  of $\alpha_s^{(5)}(M_b)$ from $\Lambda$ based on
Eq.~(\ref{eq:alp}).
\begin{verbatim}
In[4]:= lamim = LamImpl[asMz/.NumDef,Mz/.NumDef,5,3]

Out[4]= 0.208348

In[5]:= AlphasLam[lamim,Mb/.NumDef,5,3]

Out[5]= 0.216444
\end{verbatim}
Comment: evaluation of  $\alpha_s^{(5)}(M_b)$ from $\alpha_s^{(5)}(M_Z)$ based
on Eq.~(\ref{eq:defbeta}).
\begin{verbatim}
In[6]:= AlphasExact[asMz/.NumDef,Mz/.NumDef,Mb/.NumDef,5,3]

Out[6]= 0.216712
\end{verbatim}
Rounding to three significant digits leads to a difference of $\pm1$ in the
last digit. Considering the direct integration of~(\ref{eq:defbeta}) as the
most precise one we can conclude $\alpha_s^{(5)}(M_b)=0.217$
assuming three-loop accuracy.

In Tab.~\ref{tab:alsMb} the influence of the number of loops is studied
in the evaluation of $\alpha_s^{(5)}$ at the scale $M_b$ and the
(hypothetical) scale 1~GeV using $\alpha_s^{(5)}(M_Z)$ as
input. (For the latter only the function {\tt AlphasExact[]} is used for the
computation.)
It can be seen that the inclusion of $\beta_1$ leads to a significant
jump in $\alpha_s^{(5)}(M_b)$ whereas the effect of the three- and four-loop
coefficients, i.e. $\beta_2$ and $\beta_3$,
is only marginal. Their influence is more pronounced for
$\mu=1$~GeV.

\begin{table}[t]
  \begin{center}
    \begin{tabular}{c|c|l|l|l|l}
      \hline
      number   & highest coefficient & 
      \multicolumn{3}{c|}{$\alpha_s^{(5)}(M_b)$} &
      $\alpha_s^{(5)}(1~\mbox{GeV})$ \\
      of loops & in $\beta$ function &
      (a) & (b) & (c) & (c)\\
      \hline
      1 & $\beta_0$ & 0.2059 & 0.2059 & 0.2059 & 0.3369 \\
      2 & $\beta_1$ & 0.2123 & 0.2173 & 0.2161 & 0.3965 \\
      3 & $\beta_2$ & 0.2166 & 0.2164 & 0.2167 & 0.4029 \\
      4 & $\beta_3$ & 0.2174 & 0.2173 & 0.2169 & 0.4065 \\
      \hline
    \end{tabular}
    \caption{\label{tab:alsMb}$\alpha_s^{(5)}(M_b)$ and
      $\alpha_s^{(5)}(1~\mbox{GeV})$ computed from 
      $\alpha_s^{(5)}(M_Z)$ using different number of loops for the running.
      For the computation
      {\tt LamExpl[]} (a),
      {\tt LamImpl[]} (b) (both in combination with {\tt AlphasLam[]}) and
      {\tt AlphasExact[] (c)} is used.
      }
  \end{center}
\end{table}


\section{Quark masses in the $\overline{\rm MS}$ and on-shell scheme}
\label{sec:masses}

In the $\overline{\rm MS}$ scheme the running of the quark masses is governed
by the function $\gamma_{m}(\alpha_s)$
\begin{eqnarray}
  \mu^2\,\frac{d}{d\mu^2}m^{(n_f)}(\mu)
  &=&
  m^{(n_f)}(\mu)\,\gamma_m^{(n_f)}\left(\alpha_s^{(n_f)}\right) 
  \,\,=\,\,
  -m^{(n_f)}(\mu)\,\sum_{i\ge0} \gamma_{m,i}^{(n_f)}
  \left(\frac{\alpha_s^{(n_f)}(\mu)}{\pi}\right)^{i+1}
  \label{eq:defgamma}
  \,,
\end{eqnarray}
where the coefficients $\gamma_{m,i}$ are known up to the four-loop
order~\cite{gm2,Tar82,Larin:massQCD,Che97LarRitVer972}
\begin{eqnarray}
  \gamma_{m,0}^{(n_f)} &=& 1\,,
  \nonumber\\
  \gamma_{m,1}^{(n_f)} &=& \frac{1}{16}\left[ \frac{202}{3}
    - \frac{20}{9} n_f \right]\,,       
  \nonumber \\
  \gamma_{m,2}^{(n_f)} &=& \frac{1}{64} \left[1249+\left( - \frac{2216}{27} 
      - \frac{160}{3}\zeta_3 \right)n_f 
    - \frac{140}{81} n_f^2 \right]\,,
  \nonumber \\
  \gamma_{m,3}^{(n_f)} &=& \frac{1}{256} \left[ 
    \frac{4603055}{162} + \frac{135680}{27}\zeta_3 - 8800\zeta_5
    +\left(- \frac{91723}{27} - \frac{34192}{9}\zeta_3 
      + 880\zeta_4 
    \right.\right.
  \nonumber \\
  &&{}+ \left.\left.
      \frac{18400}{9}\zeta_5 \right) n_f
    +\left( \frac{5242}{243} + \frac{800}{9}\zeta_3 
      - \frac{160}{3}\zeta_4 \right) n_f^2
  \right.
  \nonumber \\&& \left.\mbox{}
    +\left(- \frac{332}{243} + \frac{64}{27}\zeta_3 \right) n_f^3 \right]
  \,,
\end{eqnarray}
with $\zeta_3\approx1.202\,057$,
$\zeta_4=\pi^4/90$ and $\zeta_5\approx1.036\,928$.
In analogy to~(\ref{eq:bidef}) we define
\begin{eqnarray}
  c_i^{(n_f)} &=& \frac{\gamma_{m,i}^{(n_f)}}{\beta_0^{(n_f)}}
  \,.
\end{eqnarray}

Combining Eqs.~(\ref{eq:defbeta}) and~(\ref{eq:defgamma}) leads to a
differential equation for $m(\mu)$ as a function of $\alpha_s(\mu)$.
It has the solution~\cite{GraBroGraSch90}
\begin{equation}
  \frac{m(\mu)}{m(\mu_0)}=
  \frac{c(\alpha_s(\mu)/\pi)}{c(\alpha_s(\mu_0)/\pi)}
  \,,
  \label{eq:mrun}
\end{equation}
with~\cite{Che97LarRitVer972}
\begin{eqnarray}
  c(x)&=&x^{c_0}\left\{1+(c_1-b_1c_0)x
    +\frac{1}{2}\left[(c_1-b_1c_0)^2+c_2-b_1c_1+b_1^2c_0-b_2c_0\right]x^2
  \right.  
  \nonumber\\
  &&{}+\left[\frac{1}{6}(c_1-b_1c_0)^3
    +\frac{1}{2}(c_1-b_1c_0)\left(c_2-b_1c_1+b_1^2c_0-b_2c_0\right)\right.
  \nonumber\\
  &&{}+\left.\left.
      \frac{1}{3}\left(c_3-b_1c_2+b_1^2c_1-b_2c_1-b_1^3c_0+2b_1b_2c_0-b_3c_0
      \right)
    \right]x^3\right\}
  \,,
  \label{eq:mas}
\end{eqnarray}
where terms of ${\cal O}(x^4)$ have been neglected.  For a given mass, $m$,
at scale $\mu_0$ and $\alpha_s(\mu_0)$ the scale invariant mass
$\mu_m=m(\mu_m)$ can be obtained from Eq.~(\ref{eq:mrun}) by
iteration.
Note the appearance of $\alpha_s(\mu)$ on the r.h.s. of~(\ref{eq:mrun}).
Thus for the computation of $\mu_m$ it is convenient to use in a first step
$\alpha_s(\mu_0)$ in combination with Eq.~(\ref{eq:alp}) to determine
$\Lambda$. Afterwards Eq.~(\ref{eq:alp}) is used again for the calculation of
$\alpha_s(\mu)$ which is inserted in~(\ref{eq:mrun}) before the iteration.

From Eq.~(\ref{eq:mrun}) it appears natural to define the 
mass~\cite{FlorNarRaf79}
\begin{eqnarray}
  \hat{m} & \equiv & \frac{m(\mu)}{c(\alpha_s(\mu)/\pi)}
  \,,
  \label{eq:mhat}
\end{eqnarray}
which is often used in the context of lattice calculations.
By construction the mass $\hat{m}$ is scale independent. It is furthermore
scheme independent (as far as mass-independent schemes are concerned). This
can be seen by considering the r.h.s. of~(\ref{eq:mhat}) in the limit
$\mu\to\infty$
\begin{eqnarray}
  \hat{m} & = & \lim_{\mu\to\infty} m(\mu)
  \left(\frac{\alpha_s(\mu)}{\pi}\right)^{-\frac{\gamma_{m,0}}{\beta_0}}
  \,,
  \label{eq:mhat2}
\end{eqnarray}
and by recalling
the fact that the coefficients $\beta_0$ and $\gamma_{m,0}$ are scheme
independent. In the following we will refer to $\hat{m}$ as
renormalization group invariant mass.

In the following we want to provide the relations between the 
$\overline{\rm MS}$ and the on-shell mass.
Whereas the coefficient of order $\alpha_s^2$ has been available
since quite some time~\cite{GraBroGraSch90,FleJegTarVer99}
only recently the three-loop
result could be obtained~\cite{CheSte99,MelRit99}.
In~\cite{CheSte99} an asymptotic expansion in combination with conformal
mapping and Pad\'e approximation has been used in order to obtain a numerical
result for the $\overline{\rm MS}$--on-shell conversion formula.
The numerical results of~\cite{CheSte99} are in perfect agreement with the 
subsequent analytical calculation of~\cite{MelRit99} (cf. Tab.~\ref{tab:nl}).
For a given on-shell
mass the $\overline{\rm MS}$ quantity can be computed with the help of
\begin{eqnarray}
  \frac{m(\mu)}{M} &=&
  1 
  + \frac{\alpha_s^{(n_f)}(\mu)}{\pi}
  \left[-\frac{4}{3}-\lmM\right]
  + \left(\frac{\alpha_s^{(n_f)}(\mu)}{\pi}\right)^2 
  \Bigg[
  -\frac{3019}{288}
  - 2\zeta_2 
  - \frac{2}{3}\zeta_2\ln2
  + \frac{1}{6}\zeta_3
  \nonumber\\&&\mbox{}
  - \frac{445}{72}\lmM
  - \frac{19}{24}\lmM^2
  + \left(\frac{71}{144} 
    + \frac{1}{3}\zeta_2
    + \frac{13}{36}\lmM
    + \frac{1}{12}\lmM^2
  \right)n_l
  - \frac{4}{3}\sum_{1\le i\le n_l} \Delta\left(\frac{\smM_i}{M}\right)
  \Bigg]
  \nonumber\\&&\mbox{}
  + \left(\frac{\alpha_s^{(n_f)}(\mu)}{\pi}\right)^3
  \Bigg[z_m^{(3)}(M)
  + \left(
    -\frac{165635}{2592} 
    - \frac{25}{3}\zeta_2
    - \frac{25}{9}\zeta_2\ln2
    + \frac{55}{36}\zeta_3
  \right)  \lmM
  \nonumber\\&&\mbox{}
  -\frac{11779}{864}\lmM^2
  -\frac{475}{432}\lmM^3
  + n_l\left(
    \left(
      \frac{10051}{1296} 
      + \frac{37}{18}\zeta_2
      + \frac{2}{9}\zeta_2\ln2
      + \frac{7}{9}\zeta_3
    \right)\lmM
    +\frac{911}{432}\lmM^2
    \right.\nonumber\\&&\left.\mbox{}
    +\frac{11}{54}\lmM^3
  \right)
  + n_l^2\left(
    \left(
      -\frac{89}{648} 
      -\frac{1}{9}\zeta_2
    \right)\lmM^2
    -\frac{13}{216}\lmM^2
    -\frac{1}{108}\lmM^3
  \right)
  \Bigg]
  \,,
  \label{eq:zmlog}
\end{eqnarray}
where $\zeta_2=\pi^2/6$ and $\lmM=\ln\mu^2/M^2$.
$n_l$ is the number of light quarks.
The function $\Delta(x)$ arises from the two-loop diagram with a second
fermion-loop~\cite{GraBroGraSch90}. For $0\leq x\leq1$
it is approximated within an
accuracy of 1\% by
\begin{eqnarray}
  \Delta(x) &=& \frac{\pi^2}{8}\,x - 0.597\,x^2 + 0.230\,x^3
  \,.
\end{eqnarray}
The corresponding mass effects at order $\alpha_s^3$ are not yet known.
In the argument of $\Delta(x)$ the ratio of the on-shell mass of the light
quarks, $\smM_i$, and the heavy one, $M$, appears.

The coefficients $z_m^{(3)}(M)$ can be found in
Tab.~\ref{tab:nl} for different values of $n_l$
where both the results of~\cite{CheSte99} and~\cite{MelRit99} are listed.
For completeness also the corresponding two-loop coefficients
(without the contribution from $\Delta(x)$) are given.
The analytical result for $z_m^{(3)}(M)$ reads~\cite{MelRit99}
\begin{eqnarray}
  z_m^{(3)}(M) &=&
  - \frac {9478333}{93312} 
  + \frac {55}{162}\ln^4 2 
  +\left( - \frac {644201}{6480}
    + \frac {587}{27}\ln 2 
    + \frac {44}{27} \ln^2 2 
  \right)\zeta_2
  - \frac {61}{27}\zeta_3 
  \nonumber\\&&\mbox{}
  + \frac {3475}{432} \zeta_4
  + \frac {1439}{72}\zeta_2\zeta_3
  - \frac {1975}{216}\zeta_5 
  + \frac {220}{27} a_4 
  + n_l  \left[ \frac {246643}{23328} 
    - \frac {1}{81}\ln^4 2  
  \nonumber\right.\\&&\left.\mbox{}
    +\left(
        \frac {967}{108}
      + \frac {22}{27}\ln 2 
      - \frac {4}{27} \ln^2 2 
    \right)\zeta_2
    + \frac {241}{72}\zeta_3
    - \frac {305}{108}\zeta_4
    - \frac {8}{27}a_4 
  \right]
  \nonumber\\&&\mbox{}
  + n_l^2  \left[  - \frac {2353}{23328} 
    - \frac {13}{54}\zeta_2
    - \frac {7}{54}\zeta_3 
  \right]
  \,,
  \label{eq:zm3}
\end{eqnarray}
where $a_4=\mbox{Li}_4(1/2)\approx 0.517\,479$.

{\tiny
  \begin{table}[t]
    \begin{center}
      \begin{tabular}{|l||r|r|r|r|r|r|r|r|r|} 
        \hline
        & \multicolumn{3}{c|}{$z_m(M)=m(M)/M$}
        & \multicolumn{3}{c|}{$z_m^{SI}(M)=\mu_m/M$}
        & \multicolumn{3}{c|}{$z_m^{inv}(m)=M/\mu_m$}
        \\
        \hline
        $n_l$ 
        & ${\cal O}(\alpha_s^2)$ 
        & ${\cal O}(\alpha_s^3)$~\cite{CheSte99}
        & ${\cal O}(\alpha_s^3)$~\cite{MelRit99}
        & ${\cal O}(\alpha_s^2)$ 
        & ${\cal O}(\alpha_s^3)$~\cite{CheSte99}
        & ${\cal O}(\alpha_s^3)$~\cite{MelRit99}
        & ${\cal O}(\alpha_s^2)$ 
        & ${\cal O}(\alpha_s^3)$~\cite{CheSte99}
        & ${\cal O}(\alpha_s^3)$~\cite{MelRit99}
        \\
        \hline
$0$ &
$    -14.33$ & $   -202(5)$ & $-198.7$ &
$    -11.67$ & $   -170(5)$ & $-166.3$ &
$     13.44$ & $    194(5)$ & $190.6$ \\
$1$ &
$    -13.29$ & $   -176(4)$ & $-172.4$ &
$    -10.62$ & $   -146(4)$ & $-142.5$ &
$     12.40$ & $    168(4)$ & $164.6$ \\
$2$ &
$    -12.25$ & $   -150(3)$ & $-147.5$ &
$     -9.58$ & $   -123(3)$ & $-120.0$ &
$     11.36$ & $    143(3)$ & $139.9$ \\
$3$ &
$    -11.21$ & $   -126(3)$ & $-123.8$ &
$     -8.54$ & $   -101(3)$ & $-98.76$ &
$     10.32$ & $    119(3)$ & $116.5$ \\
$4$ &
$    -10.17$ & $   -103(2)$ & $-101.5$ &
$     -7.50$ & $    -81(2)$ & $-78.86$ &
$      9.28$ & $     96(2)$ & $94.42$ \\
$5$ &
$     -9.13$ & $    -82(2)$ & $-80.40$ &
$     -6.46$ & $    -62(2)$ & $-60.27$ &
$      8.24$ & $     75(2)$ & $73.64$ \\
        \hline
      \end{tabular}
      \caption{\label{tab:nl}
        Two- and three-loop coefficients of the relation between on-shell and
        $\overline{\rm MS}$ mass.
        The choice $\mu^2=M^2$, respectively, $\mu^2=m^2$
        has been adopted.
        }
    \end{center}
  \end{table}
  }

Iterating~(\ref{eq:zmlog}) leads to a relation between the
scale-invariant mass, $\mu_m=m(\mu_m)$, and the on-shell mass
\begin{eqnarray}
  \frac{\mu_m}{M} &=&
  1 
  + \frac{\alpha_s^{(n_f)}(M)}{\pi}
  \left[-\frac{4}{3}\right]
  + \left(\frac{\alpha_s^{(n_f)}(M)}{\pi}\right)^2
  \Bigg[
  -\frac{2251}{288}
  - 2\zeta_2 
  - \frac{2}{3}\zeta_2\ln2
  + \frac{1}{6}\zeta_3
  \nonumber\\&&\mbox{}
  + n_l\left(
    \frac{71}{144} 
    + \frac{1}{3}\zeta_2
  \right)
  - \frac{4}{3}\sum_{1\le i\le n_l} \Delta\left(\frac{\smM_i}{M}\right)
  \Bigg]
  + \left(\frac{\alpha_s^{(n_f)}(M)}{\pi}\right)^3
  z_m^{SI,(3)}(M)
  \,.
  \label{eq:zmzm}
\end{eqnarray}
Inverting Eq.~(\ref{eq:zmlog}) leads to
\begin{eqnarray}
  \frac{M}{\mu_m} &=&
  1 
  + \frac{\alpha_s^{(n_f)}(\mu_m)}{\pi} \frac{4}{3}
  + \left(\frac{\alpha_s^{(n_f)}(\mu_m)}{\pi}\right)^2
  \Bigg[
  \frac{307}{32}
  + 2\zeta_2 
  + \frac{2}{3}\zeta_2\ln2
  - \frac{1}{6}\zeta_3
  \nonumber\\&&\mbox{}
  + n_l\left(
    - \frac{71}{144} 
    - \frac{1}{3}\zeta_2
  \right)
  + \frac{4}{3}\sum_{1\le i\le n_l} \Delta\left(\frac{m_i}{\mu_m}\right)
  \Bigg]
  + \left(\frac{\alpha_s^{(n_f)}(\mu_m)}{\pi}\right)^3
  z_m^{inv,(3)}(\mu_m)
  \,,
  \nonumber\\
  \label{eq:zminv}
\end{eqnarray}
where for convenience $\mu^2=m^2$ has been chosen.
The numerical values of the coefficients $z_m^{SI}$ and 
$z_m^{inv}$ can also be found in Tab.~\ref{tab:nl}.
Their analytic expressions are easily obtained from Eqs.~(\ref{eq:zmlog})
and~(\ref{eq:zm3}).

Eq.~(\ref{eq:zminv}) can be used to
compute the on-shell quark mass if the corresponding mass in the 
$\overline{\rm MS}$ scheme is provided.
In order to avoid large logarithms
it is suggestive to use in a first step the renormalization group
equation~(\ref{eq:mrun}) and evaluate $\mu_m$.
In a second step Eq.~(\ref{eq:zminv}) is used for $\mu=\mu_m$.
Also in the case when the on-shell mass is given
it is advantageous to use Eq.~(\ref{eq:zminv}) for the computation of the 
$\overline{\rm MS}$ mass. The reason is that Eq.~(\ref{eq:zmlog})
contains contributions from the ill-defined pole mass of the light quarks
like, e.g., the strange quark.
In the case of the top quark it is safe to use~(\ref{eq:zmlog}) as in general
the contributions for the charm and strange quark masses can be neglected.

Concerning the determination of the quark masses a crucial role is played by
lattice calculations. There it is not possible to use directly the
$\overline{\rm MS}$ scheme as it is tightly connected to dimensional
regularization.
Rather one has to use a prescription which is based on the so-called
momentum subtraction scheme. In general these schemes have the disadvantages
that they are not mass independent.
Recently, however, a mass definition based on momentum subtraction --- the
regularization invariant (RI) mass --- has been proposed which enjoys this
feature~\cite{Mar95}.
In~\cite{CheRet99} the relation to the $\overline{\rm MS}$ mass
has been evaluated to three-loop accuracy. It reads:
\begin{eqnarray}
  \frac{m^{(n_f)}(\mu)}{m^{RI}(\mu)} &=&
  1 + \frac{\alpha_s^{(n_f)}(\mu)}{\pi}\left[ -\frac{4}{3} \right]
  +\left(\frac{\alpha_s^{(n_f)}(\mu)}{\pi}\right)^2\Bigg[
  - \frac{995}{72} 
  + \frac{19}{6}\zeta_3
  + \frac{89}{144} n_f
  \Bigg]
  \nonumber\\&&\mbox{}
  +\left(\frac{\alpha_s^{(n_f)}(\mu)}{\pi}\right)^3\Bigg[
  - \frac{6663911}{41472}
  + \frac{408007}{6912}\zeta_3
  - \frac{185}{36}\zeta_5
  + \left( \frac{118325}{7776}
    + \frac{5}{12}\zeta_4
    \right.\nonumber\\&&\left.\mbox{}
    - \frac{617}{216}\zeta_3
  \right)n_f
  + \left(
    - \frac{4459}{23328}
    - \frac{1}{54}\zeta_3
  \right) n_f^2
  \Bigg]
  \,.
  \label{eq:RI2MS}
\end{eqnarray}

\begin{table}[t]
  \begin{center}
    \begin{tabular}{c|l|l|l}
      \hline
      number of loops & $m_b^{(5)}(M_Z)$ & $M_b$ & $m_b^{RI}(M_Z)$ \\
                      & (GeV) & (GeV) & (GeV) \\
      \hline
      1 & 2.903 & 4.332 & 3.048 \\
      2 & 2.715 & 4.545 & 2.885 \\
      3 & 2.696 & 4.692 & 2.872 \\
      4 & 2.693 & --- & --- \\
      \hline
    \end{tabular}
    \caption{\label{tab:mbrun}Computation of $m_b^{(5)}(M_Z)$, $M_b$ and
      $m^{RI}(M_Z)$ from
      $\mu_b=m_b^{(5)}(\mu_b)=3.97$~GeV for different number of loops.
      }
  \end{center}
\end{table}

Let us at this point consider an explicit example. For a given
mass $\mu_b=m_b^{(5)}(\mu_b)=3.97$~GeV five-flavour running
(cf. Eqs.~(\ref{eq:mrun}) and~(\ref{eq:mas}))
is used in order to obtain
$m_b^{(5)}(M_Z)$ and the on-shell mass is computed with the help of
Eq.~(\ref{eq:zminv}).
Furthermore the value for $m^{RI}(M_Z)$ is evaluated.
In Tab.~\ref{tab:mbrun} the results are listed for different
number of loops.

At the end of this section we want to summarize the different mass definitions
introduced in this section in the following table:

\begin{center}
  \begin{tabular}{|l|l|}
    \hline
    $M$          & on-shell mass \\
    $m(\mu)$   & $\overline{\rm MS}$ mass \\
    $\mu_m$      & scale invariant mass \\
    $\hat{m}$  & renormalization group invariant mass \\
    $m^{RI}$     & regularization invariant mass \\
    \hline
  \end{tabular}  
\end{center}


\section{Decoupling at flavour thresholds}
\label{sec:dec}

In MS-like renormalization schemes, the Appelquist-Carazzone decoupling
theorem~\cite{AppCar75} does not in general apply to quantities that do not
represent physical observables, such as beta functions or coupling constants, 
i.e., quarks with masses much larger than the considered energy scale
do not automatically decouple.
The standard procedure to circumvent this problem is to render decoupling
explicit by using the language of effective field theory.
The formulae presented below are valid for QCD with $n_l=n_f-1$ massless quark
flavours and one heavy flavour $h$, with mass $m_h$ which is supposed to be
much larger than the energy scale.
Then, one constructs an effective $n_l$-flavour theory by requiring 
consistency with the full $n_f$-flavour theory at 
an energy scale comparable to $m_h$, 
the heavy-quark threshold
$\mu^{(n_f)}={\cal O}(m_h)$.
This leads to a nontrivial matching condition between the
couplings and light masses, $m_q$, of the two theories.
Although, $\alpha_s^{(n_l)}(m_h)=\alpha_s^{(n_f)}(m_h)$
and $m_q^{(n_l)}(m_h)=m_q^{(n_f)}(m_h)$
at leading and
next-to-leading order, this relation does not generally hold at higher orders
in the $\overline{\rm MS}$ scheme.
At ${\cal O}(\alpha_s^2)$ the corresponding correction terms have been
computed in~\cite{Wei80,BerWet82,LarRitVer95}.

The connection between the strong coupling constant in the effective
and the full theory is given by
\begin{eqnarray}
  \alpha_s^{(nf-1)}(\mu) &=& \zeta_g^2 \alpha_s^{(n_f)}(\mu)
  \,,
  \label{eq:decdef}
\end{eqnarray}
where $\zeta_g$ is known up to the three-loop
order~\cite{CheKniSte97,CheKniSte98}:
\begin{eqnarray}
  \left(\zeta_g^{MS}\right)^2 &=& 1+
  \frac{\alpha_s^{(n_f)}(\mu)}{\pi}
  \left(
    -\frac{1}{6}\ln\frac{\mu^2}{m_h^2}
  \right)
  +\left(\frac{\alpha_s^{(n_f)}(\mu)}{\pi}\right)^2
  \left(
    \frac{11}{72} 
    -\frac{11}{24}\ln\frac{\mu^2}{m_h^2}
    +\frac{1}{36}\ln^2\frac{\mu^2}{m_h^2}
  \right)
  \nonumber\\
  &&{}+\left(\frac{\alpha_s^{(n_f)}(\mu)}{\pi}\right)^3
  \left[
    \frac{564731}{124416} 
    -\frac{82043}{27648}\zeta_3
    -\frac{955}{576}\ln\frac{\mu^2}{m_h^2}
    +\frac{53}{576}\ln^2\frac{\mu^2}{m_h^2}
  \right.
  \nonumber\\
  &&{}
  \left.
    -\frac{1}{216}\ln^3\frac{\mu^2}{m_h^2} 
    +n_l\left(
      -\frac{2633}{31104}
      +\frac{67}{576}\ln\frac{\mu^2}{m_h^2} 
      -\frac{1}{36}\ln^2\frac{\mu^2}{m_h^2}
    \right)
  \right]
  \,.
  \label{eq:zetagMS}
\end{eqnarray}
In this equation the $\overline{\rm MS}$ mass $m_h(\mu)$
--- indicated by the superscript
MS --- is chosen for the parameterization of
the heavy quark mass and $\mu$ represents the renormalization scale.
Often it is convenient to express $\zeta_g$ through the scale
invariant mass, denoted by $\mu_h=m_h(\mu_h)$:
\begin{eqnarray}
  \left(\zeta_g^{SI}\right)^2 &=& 1 +
  \frac{\alpha_s^{(n_f)}(\mu)}{\pi}
  \left(-\frac{1}{6}\ln\frac{\mu^2}{\mu_h^2}\right) 
  +\left(\frac{\alpha_s^{(n_f)}(\mu)}{\pi}\right)^2
  \left(\frac{1}{36}\ln^2\frac{\mu^2}{\mu_h^2}
    -\frac{19}{24}\ln\frac{\mu^2}{\mu_h^2}+\frac{11}{72}\right)
  \nonumber\\&&\mbox{}
  +\left(\frac{\alpha_s^{(n_f)}(\mu)}{\pi}\right)^3
  \left[-\frac{1}{216}\ln^3\frac{\mu^2}{\mu_h^2}
    -\frac{131}{576}\ln^2\frac{\mu^2}{\mu_h^2}
    +\frac{1}{1728}\ln\frac{\mu^2}{\mu_h^2}(-6793+281\,n_l)
  \right.\nonumber\\
  &&\left.
    -\frac{82043}{27648}\zeta_3+\frac{564731}{124416}
    -\frac{2633}{31104}n_l
  \right]
  \,.
  \label{eq:zetagSI}
\end{eqnarray}
Transforming the heavy quark mass into the on-shell scheme leads to
\begin{eqnarray}
  \left(\zeta_g^{OS}\right)^2 &=& 1 +
  \frac{\alpha_s^{(n_f)}(\mu)}{\pi}
  \left(
    -\frac{1}{6}\ln\frac{\mu^2}{M_h^2}
  \right)
  +\left(\frac{\alpha_s^{(n_f)}(\mu)}{\pi}\right)^2
  \left(
    -\frac{7}{24} 
    -\frac{19}{24}\ln\frac{\mu^2}{M_h^2}
    +\frac{1}{36}\ln^2\frac{\mu^2}{M_h^2}
  \right)
  \nonumber\\
  &&{}+\left(\frac{\alpha_s^{(n_f)}(\mu)}{\pi}\right)^3
  \left[
    -\frac{58933}{124416}
    -\frac{2}{3}\zeta_2\left(1+\frac{1}{3}\ln2\right)
    -\frac{80507}{27648}\zeta_3
    -\frac{8521}{1728}\ln\frac{\mu^2}{M_h^2}
  \right.\nonumber\\
  &&{}-\left.
    \frac{131}{576}\ln^2\frac{\mu^2}{M_h^2}
    -\frac{1}{216}\ln^3\frac{\mu^2}{M_h^2} 
    +n_l\left(
      \frac{2479}{31104}
      +\frac{\zeta_2}{9}
      +\frac{409}{1728}\ln\frac{\mu^2}{M_h^2} 
    \right)
  \right]
  \,.
  \label{eq:zetagOS}
\end{eqnarray}

In practical applications also the inverted formulae are needed
which read for Eqs.~(\ref{eq:zetagMS}),~(\ref{eq:zetagSI})
and~(\ref{eq:zetagOS}):
\begin{eqnarray}
  \frac{1}{\left(\zeta_g^{MS}\right)^2} &=&
  1+
  \frac{\alpha_s^{(n_l)}(\mu)}{\pi}
  \left(
    \frac{1}{6}\ln\frac{\mu^2}{m_h^2}
  \right)
  +\left(\frac{\alpha_s^{(n_l)}(\mu)}{\pi}\right)^2
  \left(
    -\frac{11}{72} 
    +\frac{11}{24}\ln\frac{\mu^2}{m_h^2}
    +\frac{1}{36}\ln^2\frac{\mu^2}{m_h^2}
  \right)
  \nonumber\\
  &&{}+\left(\frac{\alpha_s^{(n_l)}(\mu)}{\pi}\right)^3
  \left[
    -\frac{564731}{124416} 
    +\frac{82043}{27648}\zeta_3
    +\frac{2645}{1728}\ln\frac{\mu^2}{m_h^2}
    +\frac{167}{576}\ln^2\frac{\mu^2}{m_h^2}
  \right.
  \nonumber\\
  &&{}
  \left.
    +\frac{1}{216}\ln^3\frac{\mu^2}{m_h^2} 
    +n_l\left(
       \frac{2633}{31104}
      -\frac{67}{576}\ln\frac{\mu^2}{m_h^2} 
      +\frac{1}{36}\ln^2\frac{\mu^2}{m_h^2}
    \right)
  \right]
  \,,
  \label{eq:invzetagMS}
  \\
  \frac{1}{\left(\zeta_g^{SI}\right)^2} &=&
  1+
  \frac{\alpha_s^{(n_l)}(\mu)}{\pi}
  \left(
    \frac{1}{6}\ln\frac{\mu^2}{\mu_h^2}
  \right)
  +\left(\frac{\alpha_s^{(n_l)}(\mu)}{\pi}\right)^2
  \left(
    -\frac{11}{72} 
    +\frac{19}{24}\ln\frac{\mu^2}{\mu_h^2}
    +\frac{1}{36}\ln^2\frac{\mu^2}{\mu_h^2}
  \right)
  \nonumber\\
  &&{}+\left(\frac{\alpha_s^{(n_l)}(\mu)}{\pi}\right)^3
  \left[
    -\frac{564731}{124416} 
    +\frac{82043}{27648}\zeta_3
    +\frac{2191}{576}\ln\frac{\mu^2}{\mu_h^2}
    +\frac{511}{576}\ln^2\frac{\mu^2}{\mu_h^2}
  \right.
  \nonumber\\
  &&{}
  \left.
    +\frac{1}{216}\ln^3\frac{\mu^2}{\mu_h^2} 
    +n_l\left(
      \frac{2633}{31104}
      -\frac{281}{1728}\ln\frac{\mu^2}{\mu_h^2} 
    \right)
  \right]
  \,,
  \label{eq:invzetagSI}
  \\
  \frac{1}{\left(\zeta_g^{OS}\right)^2} &=&
  1 +
  \frac{\alpha_s^{(n_l)}(\mu)}{\pi}
  \left(
    \frac{1}{6}\ln\frac{\mu^2}{M_h^2}
  \right)
  +\left(\frac{\alpha_s^{(n_l)}(\mu)}{\pi}\right)^2
  \left(
     \frac{7}{24} 
    +\frac{19}{24}\ln\frac{\mu^2}{M_h^2}
    +\frac{1}{36}\ln^2\frac{\mu^2}{M_h^2}
  \right)
  \nonumber\\
  &&{}+\left(\frac{\alpha_s^{(n_l)}(\mu)}{\pi}\right)^3
  \left[
    \frac{58933}{124416}
    +\frac{2}{3}\zeta_2\left(1+\frac{1}{3}\ln2\right)
    +\frac{80507}{27648}\zeta_3
    +\frac{8941}{1728}\ln\frac{\mu^2}{M_h^2}
  \right.\nonumber\\
  &&{}+\left.
    \frac{511}{576}\ln^2\frac{\mu^2}{M_h^2}
    +\frac{1}{216}\ln^3\frac{\mu^2}{M_h^2} 
    +n_l\left(
      -\frac{2479}{31104}
      -\frac{\zeta_2}{9}
      -\frac{409}{1728}\ln\frac{\mu^2}{M_h^2} 
    \right)
  \right]
  \,.
  \label{eq:invzetagOS}
\end{eqnarray}
The decoupling relations~(\ref{eq:zetagMS})--(\ref{eq:invzetagOS})
have to be applied whenever a flavour threshold is to be crossed.

At this point we briefly want to comment on the order of $\alpha_s$
which has to be used for the running, respectively, the decoupling
if the analysis should be consistent.
If the $\mu$ evolution of $\alpha_s^{(n_f)}(\mu)$ is to be performed at $N+1$
loops, i.e., with the highest coefficient in Eq.~(\ref{eq:defbeta})
being $\beta_N^{(n_f)}$, then consistency requires the matching
conditions to be implemented in terms of $N$-loop formulae.
Then, the residual $\mu$ dependence of physical observables will be of order
$N+2$.

As an example let us compute $\alpha_s^{(4)}(M_c)$ from
$\alpha_s^{(5)}(M_Z)=0.118$. Let us furthermore consider the on-shell
definition of the heavy quark, $M_b$, i.e. we use 
Eq.~(\ref{eq:zetagOS}) for the analysis.
For the scale $\mu$ in~(\ref{eq:decdef}) where the matching is performed we
choose $\mu_{th}=M_b$. Assuming four-loop accuracy for the beta function
and (as a consequence) three-loop accuracy for the matching
computation one would proceed as follows (see Appendix for a description of the
procedures): 
\begin{verbatim}
In[2]:= (alsmuth = AlphasExact[asMz/.NumDef,Mz/.NumDef,Mb/.NumDef,5,4])

Out[2]= 0.2169467

In[3]:= (alsmuthp = DecAsDownOS[alsmuth,Mb/.NumDef,Mb/.NumDef,4,4])

Out[3]= 0.2163396

In[4]:= (alsMc = AlphasExact[alsmuthp,Mb/.NumDef,Mc/.NumDef,4,4])

Out[4]= 0.337848
\end{verbatim}
Finally one arrives at $\alpha_s^{(4)}(M_c)=0.338$.
These steps are summarized in the function {\tt AsRunDec[]} where the
corresponding call would read
\begin{verbatim}
In[5]:= AsRunDec[asMz/.NumDef,Mz/.NumDef,Mc/.NumDef,4]

Out[5]= 0.337848
\end{verbatim}
Note that the loop-argument of the function {\tt DecAsDownOS[]}
(last argument) refers to the order used for the running, i.e.
in the considered case the ``4'' means that the three-loop relation is used
for the decoupling.
In this example the effect of the decoupling is quite small.
It is actually comparable to the uncertainty from  using different
methods for the running (cf. Tab.~\ref{tab:alsMb}).
However, one has to remember that for the matching scale the heavy quark mass
itself has been used, whence  all logarithms in Eq.~(\ref{eq:zetagOS})
vanish. A different choice would lead to a different result for
$\alpha_s^{(4)}(M_c)$. On the other hand, on general grounds, the decoupling
procedure should not depend on the choice of that scale, respectively,
the dependence should become weaker when going to higher orders.
In Tab.~\ref{tab:dec} the dependence of $\alpha_s^{(4)}(M_c)$ on the number of
loops is shown. For the matching scale $M_Z$, $M_b$ and $1$~GeV has been
chosen. 
It can be clearly seen that the four-loop analysis provides the most stable
values --- even in the case when the matching is performed at the 
a high scale like the $Z$ boson mass.
This is expected on general grounds as physical results should not depend
on the matching scale.
\begin{table}[t]
  \begin{center}
    \begin{tabular}{c|l|l|l}
      \hline
      number of loops &
      \multicolumn{3}{c}{$\alpha_s^{(4)}(M_c)$} \\
      (running) & 
      ($\mu_{th}=M_Z$) & ($\mu_{th}=M_b$) & ($\mu_{th}=1$~GeV) \\
      \hline
      1 & 0.3213 & 0.2918 & 0.2784 \\
      2 & 0.3387 & 0.3318 & 0.3222 \\
      3 & 0.3396 & 0.3364 & 0.3319 \\
      4 & 0.3399 & 0.3378 & 0.3350 \\
      \hline
    \end{tabular}
    \caption{\label{tab:dec}Computation of $\alpha_s^{(4)}(M_c)$ from
      $\alpha_s^{(5)}(M_Z)=0.118$ for different number of loops. For the
      matching scale $M_Z$, $M_b$ and 1~GeV has been chosen. 
      }
  \end{center}
\end{table}

We should mention that in case the $\overline{\rm MS}$ definition for the
heavy quark is used in a first step $m_h(\mu_{th})$ has to be evaluated. The
corresponding formulae can be found in Section~\ref{sec:masses}.
They are also implemented as {\tt Mathematica} procedures and described in
the Appendix.

In Fig.~\ref{fig:asMZasMtau} it is demonstrated that the inclusion of the
four-loop coefficient $\beta_3$ accompanied by the three-loop matching
leads to an independence of $\mu_{th}=\mu^{(5)}$ over a very broad
range~\cite{CheKniSte98}.
The plot shows the dependence of $\alpha_s^{(5)}(M_Z)$  on the matching scale
(denoted by $\mu^{(5)}$) where  $\alpha_s^{(4)}(M_\tau)$ is used as starting
point. Our procedure to get the different curves is as follows.
We first calculate $\alpha_s^{(4)}(\mu^{(5)})$ by exactly integrating
Eq.~(\ref{eq:defbeta}) with the initial condition
$\alpha_s^{(4)}(M_\tau)=0.36$, then obtain $\alpha_s^{(5)}(\mu^{(5)})$ from
Eqs.~(\ref{eq:invzetagOS}) with $M_b=4.7$~GeV, and finally compute
$\alpha_s^{(5)}(M_Z)$ with Eq.~(\ref{eq:defbeta}).
For consistency, $N$-loop evolution must be accompanied by $(N-1)$-loop 
matching, i.e.\ if we omit terms of ${\cal O}(\alpha_s^{N+2})$ on the 
right-hand side of Eq.~(\ref{eq:defbeta}), we need to discard those of
${\cal O}(\alpha_s^N)$ in Eq.~(\ref{eq:invzetagOS}) at the same time.
In Fig.~\ref{fig:asMZasMtau}, the variation of $\alpha_s^{(5)}(M_Z)$ with
$\mu^{(5)}/M_b$ 
is displayed for the various levels of accuracy, ranging from one-loop to
four-loop evolution.
For illustration, $\mu^{(5)}$ is varied   by almost two orders
of magnitude.
While the leading-order result exhibits a strong logarithmic behaviour, it
stabilizes  as we  go to higher orders.
The four-loop curve is almost flat for $\mu^{(5)}\gsim1$~GeV.
Besides the $\mu^{(5)}$ dependence of $\alpha_s^{(5)}(M_Z)$, also its absolute 
normalization is significantly affected by the higher orders.
At the central matching scale $\mu^{(5)}=M_b$, we encounter a rapid, monotonic
convergence behaviour.

\begin{figure}[t]
  \begin{center}
    \begin{tabular}{c}
      \leavevmode
      \epsfxsize=14cm
      \epsffile[90 275 463 579]{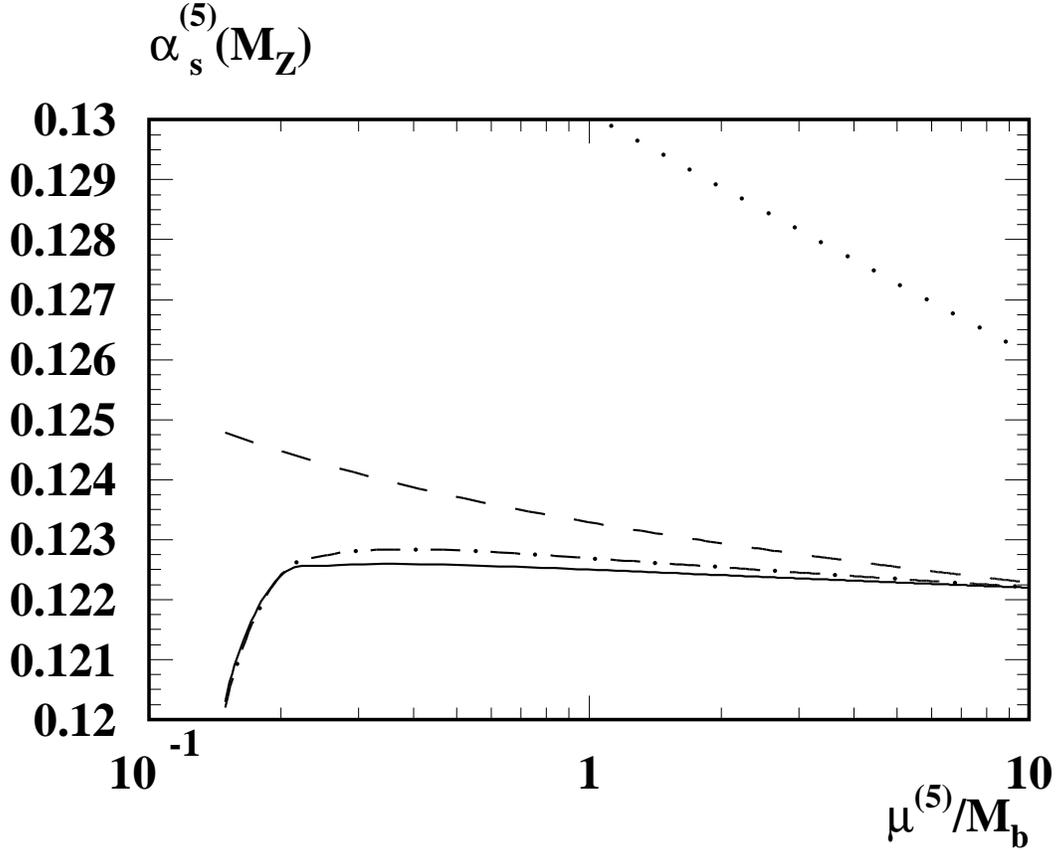}
    \end{tabular}
    \caption{\label{fig:asMZasMtau}$\mu^{(5)}$ dependence of
      $\alpha_s^{(5)}(M_Z)$ calculated from 
      $\alpha_s^{(4)}(M_\tau)=0.36$ and $M_b=4.7$~GeV using
      Eq.~(\ref{eq:defbeta}) 
      at one (dotted), two (dashed), three (dot-dashed), and four (solid)
      loops in connection with Eq.~(\ref{eq:invzetagOS}) at the 
      respective orders.}
  \end{center}
\end{figure}

Fig.~\ref{fig:asMZasMtau} can immediately be reproduced with the help of
the procedure {\tt AlL2AlH[]} described in the Appendix.

Up to now only the decoupling of the coupling constant has been
considered. However, also the (relatively) lighter quark masses
undergo a decoupling procedure when crossing a flavour threshold. If
we define the connection between the quark mass in the effective and
full theory through
\begin{eqnarray}
  m_q^{(n_f-1)} &=& \zeta_m m_q^{(n_f)}
  \,,
  \label{eq:decdefmass}
\end{eqnarray}
the decoupling constant $\zeta_m$ is given by~\cite{CheKniSte98}
\begin{eqnarray}
  \zeta_m^{MS} &=&
  1
  +\left(\frac{\alpha_s^{(n_f)}(\mu)}{\pi}\right)^2
  \left(\frac{89}{432} 
    -\frac{5}{36}\ln\frac{\mu^2}{m_h^2}
    +\frac{1}{12}\ln^2\frac{\mu^2}{m_h^2}\right)
  +\left(\frac{\alpha_s^{(n_f)}(\mu)}{\pi}\right)^3
  \left[\frac{2951}{2916} 
  \right.
  \nonumber\\
  &&\left.\mbox{}
    -\frac{407}{864}\zeta_3
   +\frac{5}{4}\zeta_4
    -\frac{1}{36}B_4
    +\left(-\frac{311}{2592}
      -\frac{5}{6}\zeta_3\right)\ln\frac{\mu^2}{m_h^2}
    +\frac{175}{432}\ln^2\frac{\mu^2}{m_h^2}
  \right.
  \nonumber\\
  &&{}\left.\mbox{}
    +\frac{29}{216}\ln^3\frac{\mu^2}{m_h^2}
    +n_l\left(
      \frac{1327}{11664}
      -\frac{2}{27}\zeta_3
      -\frac{53}{432}\ln\frac{\mu^2}{m_h^2}
      -\frac{1}{108}\ln^3\frac{\mu^2}{m_h^2}\right)\right]
  \,,
  \label{eq:zetamMS}
\end{eqnarray}
where~\cite{Bro92}
\begin{eqnarray}
  B_4&=&16\li\left({1\over2}\right)-{13\over2}\zeta_4-4\zeta_2\ln^22
  +{2\over3}\ln^42
  \nonumber\\
  &\approx&-1.762\,800\,.
\end{eqnarray}
Note that all three quantities in Eq.~(\ref{eq:decdefmass})
depend on the renormalization scale $\mu$.

Again it turns out to be useful to consider in addition to~(\ref{eq:zetamMS})
the quantities where
the scale invariant and the on-shell mass, respectively, has
been used for the parameterization of the heavy quark:
\begin{eqnarray}
  \zeta_m^{SI} &=&
  1
  +\left(\frac{\alpha_s^{(n_f)}(\mu)}{\pi}\right)^2
  \left(\frac{89}{432} 
    -\frac{5}{36}\ln\frac{\mu^2}{\mu_h^2}
    +\frac{1}{12}\ln^2\frac{\mu^2}{\mu_h^2}\right)
  +\left(\frac{\alpha_s^{(n_f)}(\mu)}{\pi}\right)^3
  \left[\frac{2951}{2916} 
  \right.
  \nonumber\\
  &&\left.\mbox{}
    -\frac{407}{864}\zeta_3
   +\frac{5}{4}\zeta_4
    -\frac{1}{36}B_4
    +\left(-\frac{1031}{2592}
      -\frac{5}{6}\zeta_3\right)\ln\frac{\mu^2}{\mu_h^2}
    +\frac{319}{432}\ln^2\frac{\mu^2}{\mu_h^2}
  \right.
  \nonumber\\
  &&{}\left.\mbox{}
    +\frac{29}{216}\ln^3\frac{\mu^2}{\mu_h^2}
    +n_l\left(
      \frac{1327}{11664}
      -\frac{2}{27}\zeta_3
      -\frac{53}{432}\ln\frac{\mu^2}{\mu_h^2}
      -\frac{1}{108}\ln^3\frac{\mu^2}{\mu_h^2}\right)\right]
  \,,
  \label{eq:zetamSI}
  \\
  \zeta_m^{OS} &=&
  1
  +\left(\frac{\alpha_s^{(n_f)}(\mu)}{\pi}\right)^2
  \left(\frac{89}{432} 
    -\frac{5}{36}\ln\frac{\mu^2}{M_h^2}
    +\frac{1}{12}\ln^2\frac{\mu^2}{M_h^2}
  \right)
  +\left(\frac{\alpha_s^{(n_f)}(\mu)}{\pi}\right)^3
  \left[
    \frac{1871}{2916} 
  \right.
  \nonumber\\
  &&
  \left.\mbox{}
    - \frac{407}{864}\zeta_3
    +\frac{5}{4}\zeta_4
    - \frac{1}{36}B_4
    +\left(\frac{121}{2592}
      - \frac{5}{6}\zeta_3\right)\ln\frac{\mu^2}{M_h^2}
    + \frac{319}{432}\ln^2\frac{\mu^2}{M_h^2}
  \right.
  \nonumber\\
  &&
  \left.\mbox{}
    + \frac{29}{216}\ln^3\frac{\mu^2}{M_h^2}
    +n_l\left(
      \frac{1327}{11664}
      - \frac{2}{27}\zeta_3
      - \frac{53}{432}\ln\frac{\mu^2}{M_h^2}
      - \frac{1}{108}\ln^3\frac{\mu^2}{M_h^2}
    \right)
  \right]
  \,.
  \label{eq:zetamOS}
\end{eqnarray}
The corresponding inverted relations read
\begin{eqnarray}
  \frac{1}{\zeta_m^{MS}} &=&
  1
  +\left(\frac{\alpha_s^{(n_l)}(\mu)}{\pi}\right)^2
  \left(-\frac{89}{432} 
    +\frac{5}{36}\ln\frac{\mu^2}{m_h^2}
    -\frac{1}{12}\ln^2\frac{\mu^2}{m_h^2}\right)
  +\left(\frac{\alpha_s^{(n_l)}(\mu)}{\pi}\right)^3
  \left[-\frac{2951}{2916} 
  \right.
  \nonumber\\
  &&\left.\mbox{}
    +\frac{407}{864}\zeta_3
    -\frac{5}{4}\zeta_4
    +\frac{1}{36}B_4
    +\left(\frac{133}{2592}
      +\frac{5}{6}\zeta_3\right)\ln\frac{\mu^2}{m_h^2}
    -\frac{155}{432}\ln^2\frac{\mu^2}{m_h^2}
  \right.
  \nonumber\\
  &&{}\left.\mbox{}
    -\frac{35}{216}\ln^3\frac{\mu^2}{m_h^2}
    +n_l\left(
      -\frac{1327}{11664}
      +\frac{2}{27}\zeta_3
      +\frac{53}{432}\ln\frac{\mu^2}{m_h^2}
      +\frac{1}{108}\ln^3\frac{\mu^2}{m_h^2}\right)\right]
  \,,
  \label{eq:invzetamMS}
  \\
  \frac{1}{\zeta_m^{SI}} &=&
  1
  +\left(\frac{\alpha_s^{(n_l)}(\mu)}{\pi}\right)^2
  \left(-\frac{89}{432} 
    +\frac{5}{36}\ln\frac{\mu^2}{\mu_h^2}
    -\frac{1}{12}\ln^2\frac{\mu^2}{\mu_h^2}\right)
  +\left(\frac{\alpha_s^{(n_l)}(\mu)}{\pi}\right)^3
  \left[-\frac{2951}{2916} 
  \right.
  \nonumber\\
  &&\left.\mbox{}
    +\frac{407}{864}\zeta_3
    -\frac{5}{4}\zeta_4
    +\frac{1}{36}B_4
    +\left(\frac{853}{2592}
      +\frac{5}{6}\zeta_3\right)\ln\frac{\mu^2}{\mu_h^2}
    -\frac{299}{432}\ln^2\frac{\mu^2}{\mu_h^2}
  \right.
  \nonumber\\
  &&{}\left.\mbox{}
    -\frac{35}{216}\ln^3\frac{\mu^2}{\mu_h^2}
    +n_l\left(
      -\frac{1327}{11664}
      +\frac{2}{27}\zeta_3
      +\frac{53}{432}\ln\frac{\mu^2}{\mu_h^2}
      +\frac{1}{108}\ln^3\frac{\mu^2}{\mu_h^2}\right)\right]
  \,,
  \label{eq:invzetamSI}
  \\
  \frac{1}{\zeta_m^{OS}} &=&
  1
  +\left(\frac{\alpha_s^{(n_l)}(\mu)}{\pi}\right)^2
  \left(-\frac{89}{432} 
    +\frac{5}{36}\ln\frac{\mu^2}{M_h^2}
    -\frac{1}{12}\ln^2\frac{\mu^2}{M_h^2}
  \right)
  +\left(\frac{\alpha_s^{(n_l)}(\mu)}{\pi}\right)^3
  \left[
    -\frac{1871}{2916} 
  \right.
  \nonumber\\
  &&
  \left.\mbox{}
    +\frac{407}{864}\zeta_3
    -\frac{5}{4}\zeta_4
    + \frac{1}{36}B_4
    +\left(-\frac{299}{2592}
      + \frac{5}{6}\zeta_3\right)\ln\frac{\mu^2}{M_h^2}
    - \frac{299}{432}\ln^2\frac{\mu^2}{M_h^2}
  \right.
  \nonumber\\
  &&
  \left.\mbox{}
    - \frac{35}{216}\ln^3\frac{\mu^2}{M_h^2}
    +n_l\left(
      - \frac{1327}{11664}
      + \frac{2}{27}\zeta_3
      + \frac{53}{432}\ln\frac{\mu^2}{M_h^2}
      + \frac{1}{108}\ln^3\frac{\mu^2}{M_h^2}
    \right)
  \right]
  \,.
  \label{eq:invzetamOS}
\end{eqnarray}

As an example we compute $m_c^{(5)}(M_Z)$ for different number of loops and
different matching points $\mu_{th}$.
The results can be found in Tab.~\ref{tab:mcrundec} where 
$\mu_{th}=M_Z,~M_b$, and 1~GeV has been chosen.
It can clearly be seen that the four-loop result provides the most stable
values for $m_c^{(5)}(M_Z)$.

\begin{table}[t]
  \begin{center}
    \begin{tabular}{c|l|l|l}
      \hline
      number of loops &
      \multicolumn{3}{c}{$m_c^{(5)}(M_Z)$} \\
      (running) & 
      ($\mu_{th}=M_Z$) & ($\mu_{th}=M_b$) & ($\mu_{th}=1$~GeV) \\
      \hline
      1 & 0.6968 & 0.7174 & 0.7207\\
      2 & 0.6025 & 0.6051 & 0.6236\\
      3 & 0.5846 & 0.5852 & 0.5984\\
      4 & 0.5798 & 0.5801 & 0.5783\\
      \hline
    \end{tabular}
    \caption{\label{tab:mcrundec}Computation of $m_c^{(5)}(M_Z)$ from
      $\mu_c=m_c(\mu_c)=1.2$~GeV for different number of loops.
      For the matching scale $M_Z$, $M_b$ and 1~GeV has been chosen.
      }
  \end{center}
\end{table}

\begin{figure}[t]
  \begin{center}
    \begin{tabular}{c}
      \leavevmode
      \epsfxsize=14cm
      \epsffile[90 275 463 579]{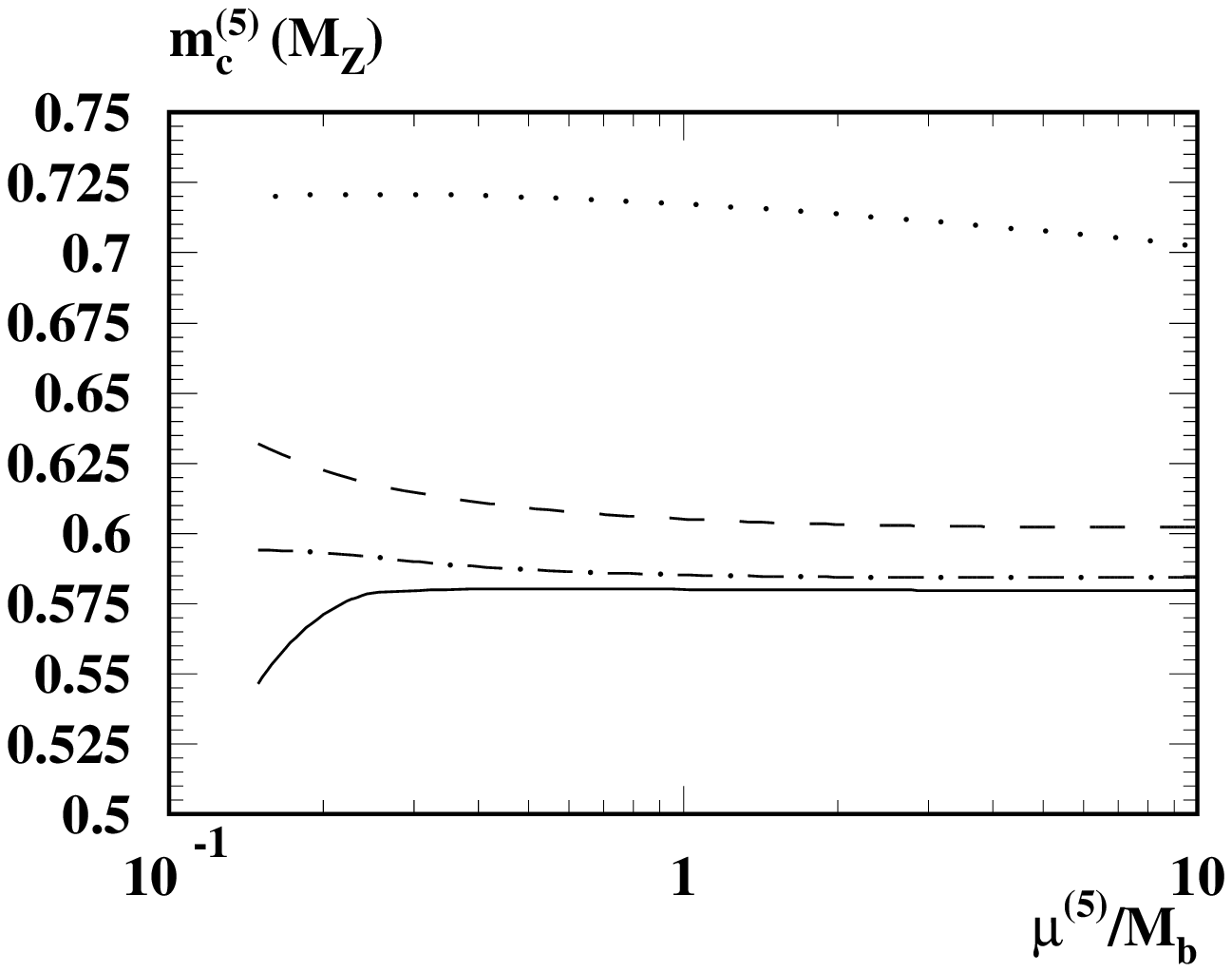}
    \end{tabular}
    \caption{\label{fig:mcMZ}$\mu^{(5)}$ dependence of
      $m_c^{(5)}(M_Z)$ calculated from 
      $\mu_c=m_c^{(4)}(\mu_c)=1.2$~GeV, $M_b=4.7$~GeV and
      $\alpha_s^{(5)}(M_Z)$ using
      Eq.~(\ref{eq:defgamma}) 
      at one (dotted), two (dashed), three (dot-dashed), and four (solid)
      loops in connection with Eq.~(\ref{eq:invzetamOS}) at the 
      respective orders.}
  \end{center}
\end{figure}

A similar analysis as in Fig.~\ref{fig:asMZasMtau} may be performed for the
light-quark masses as well.
For illustration, let us investigate how the $\mu^{(5)}$ dependence of the
relation between $\mu_c=m_c^{(4)}(\mu_c)$ and $m_c^{(5)}(M_Z)$ changes under
the inclusion of higher orders in evolution and matching.
As typical input parameters, we choose $\mu_c=1.2$~GeV, $M_b=4.7$~GeV, and
$\alpha_s^{(5)}(M_Z)=0.118$.
We first evolve $m_c^{(4)}(\mu)$ from $\mu=\mu_c$ to 
$\mu=\mu_{th}=\mu^{(5)}$ via
Eq.~(\ref{eq:mas}), then obtain $m_c^{(5)}(\mu^{(5)})$ from
Eqs.~(\ref{eq:invzetamOS}),
and finally evolve $m_c^{(5)}(\mu)$ from
$\mu=\mu^{(5)}$ to $\mu=M_Z$ via Eq.~(\ref{eq:mas}).
In all steps, $\alpha_s^{(n_f)}(\mu)$ is evaluated with the same values of
$n_f$ and $\mu$ as $m_c^{(n_f)}(\mu)$.
In Fig.~\ref{fig:mcMZ}, we show the resulting values of $m_c^{(5)}(M_Z)$ 
corresponding to $N$-loop evolution with $(N-1)$-loop matching for 
$N=1,\ldots,4$.
Similarly to Fig.~\ref{fig:asMZasMtau}, we observe a rapid, monotonic
convergence 
behaviour at the central matching scale $\mu^{(5)}=M_b$.
Again, the prediction for $N=4$ is remarkably stable under the variation of
$\mu^{(5)}$ as long as $\mu^{(5)}\gsim1$~GeV.
Fig.~\ref{fig:mcMZ} can easily be reproduced with the help of the procedure
{\tt mL2mH[]} (see Appendix).

If one chooses to perform the running of $\alpha_s(\mu)$ with the help of
$\Lambda$ it is useful to have an equation at hand which relates this
parameter in the full and effective theory.
Combining Eqs.~(\ref{eq:lamexpl}),~(\ref{eq:alp}) and~(\ref{eq:zetagSI})
one obtains~\cite{CheKniSte97}
\begin{eqnarray}
  \beta_0^\prime\ln\frac{\Lambda^{\prime2}}{\Lambda^2} &=&
  (\beta_0^\prime-\beta_0)l_h+(b_1^\prime-b_1)\ln l_h
  -b_1^\prime\ln\frac{\beta_0^\prime}{\beta_0}
  \nonumber\\
  &&{}+\frac{1}{\beta_0l_h}\left[b_1(b_1^\prime-b_1)\ln l_h
    +b_1^{\prime2}-b_1^2-b_2^\prime+b_2+\frac{11}{72}\right]
  \nonumber\\
  &&{}+\frac{1}{(\beta_0l_h)^2}\left\{-\frac{b_1^2}{2}(b_1^\prime-b_1)
    \ln^2l_h+b_1[-b_1^\prime(b_1^\prime-b_1)
  \right.\nonumber\\
  &&{}+b_2^\prime-b_2-\frac{11}{72}]\ln l_h
  +\frac{1}{2}(-b_1^{\prime3}-b_1^3-b_3^\prime+b_3)
  \nonumber\\
  &&{}+\left.\vphantom{\frac{b_1^2}{2}}
    b_1^\prime(b_1^2+b_2^\prime-b_2-\frac{11}{72})
    +\frac{564731}{124416}-\frac{82043}{27648}\zeta_3-\frac{2633}{31104}n_l
  \right\}\,,
  \label{eq:declam}
\end{eqnarray}
where $l_h=\ln(\mu_h^2/\Lambda^2)$ and the primed quantities refer to the
$(n_f-1)$-flavour effective theory.
In this equation $\mu_h$ has been chosen for the matching scale
which is particularly convenient, since it eliminates the renormalization
group logarithms in~(\ref{eq:zetagSI}).
This choice is furthermore justified with the help of
Figs.~\ref{fig:asMZasMtau} and~\ref{fig:mcMZ} where it can be seen that,
in higher orders, the actual value of the matching scale
does not matter as long as it is comparable to the heavy-quark
mass.
In Eq.~(\ref{eq:declam}) the four different powers in $l_h$ correspond to the
different loop orders. Whereas at one-loop accuracy only the linear 
term in $l_h$ has to be taken into account at four-loop order also the 
$1/l_h^2$ contribution has to be considered.
Eq.~(\ref{eq:declam}) is implemented in the procedure
{\tt DecLambdaDown[]}.

For completeness we also display the inverted relation of
Eq.~(\ref{eq:declam}):
\begin{eqnarray}
  \beta_0\ln\frac{\Lambda^2}{\Lambda^{\prime2}} &=&
  (\beta_0-\beta_0^\prime)l_h^\prime+(b_1-b_1^\prime)\ln l_h^\prime
  -b_1\ln\frac{\beta_0}{\beta_0^\prime}
  \nonumber\\
  &&{}+\frac{1}{\beta_0^\prime l_h^\prime}\left[b_1^\prime(b_1-b_1^\prime)\ln
    l_h^\prime 
    +b_1^{2}-b_1^{\prime2}-b_2+b_2^\prime-\frac{11}{72}\right]
  \nonumber\\
  &&{}+\frac{1}{(\beta_0^\prime l_h^\prime)^2}\left\{-\frac{b_1^{\prime2}}{2}
    (b_1-b_1^\prime)
    \ln^2l_h^\prime+b_1^\prime[-b_1(b_1-b_1^\prime)
  \right.\nonumber\\
  &&{}+b_2-b_2^\prime+\frac{11}{72}]\ln l_h^\prime
  +\frac{1}{2}(-b_1^{3}-b_1^{\prime3}-b_3+b_3^\prime)
  \nonumber\\
  &&{}+\left.\vphantom{\frac{b_1^2}{2}}
    b_1(b_1^{\prime2}+b_2-b_2^\prime+\frac{11}{72})
    -\frac{564731}{124416}+\frac{82043}{27648}\zeta_3+\frac{2633}{31104}n_l
  \right\}
  \,,
  \label{eq:declam2}
\end{eqnarray}
with $l_h^\prime=\ln(\mu_h^2/(\Lambda^\prime)^2)$.
It is realized in the procedure {\tt DecLambdaUp[]}.

At this point we would like to mention that next to the coupling constant and
quark masses also the gauge parameter and the quark and gluon fields
obey decoupling relations. The corresponding equations and results
can be found in~\cite{CheKniSte98}.


\section{Description of the main procedures}
\label{sec:desc}

In this section we describe the procedures which are most important for the
practical applications,  namely the combined running and decoupling of the strong
coupling and the conversion of the on-shell mass to the $\overline{\rm MS}$
one and vice versa.

In {\tt RunDec.m} some masses and couplings are set to default values
which are used if they are not specified explicitly.
They are collected in the set {\tt NumDef} and read (also
the corresponding symbol used in {\tt RunDec} is given):
\begin{eqnarray}
  \begin{array}{llllll}
    {\tt Mtau}:
    &
    M_\tau = 1.777\mbox{~GeV}\,,
    &
    {\tt Mc}:
    &
    M_c = 1.6\mbox{~GeV}\,,
    &
    {\tt Mb}:
    &
    M_b = 4.7\mbox{~GeV}\,,
    \\  
    {\tt Mt}:
    &
    M_t = 175\mbox{~GeV}\,,
    &
    {\tt muc}:
    &
    \mu_c = 1.2\mbox{~GeV}\,,
    &
    {\tt mub}:
    &
    \mu_b = 3.97\mbox{~GeV}\,,
    \\
    {\tt Mz}:
    &
    M_Z = 91.18\mbox{~GeV}\,,
    &
    {\tt asMz}:
    &
    \alpha_s^{(5)}(M_Z) = 0.118 \,.
  \end{array}
  \label{eq:numdef}
\end{eqnarray}

The following procedure computes $\alpha_s^{(m)}(\mu)$
where $\alpha_s^{(n)}(\mu_0)$ is used as input parameter.
As input only $\alpha_s(\mu_0)$, $\mu_0$, $\mu$ and the number of loops
have to be specified. Both $n$ and $m$ are determined according to the values
of the quark masses given in {\tt NumDef}.
In case $n\not=m$ the heavy quarks are consistently decoupled at the heavy
quark scale itself where for the mass definition the on-shell scheme is used.

\begin{flushleft}
\begin{itemize}
\item {\tt AsRunDec}:
  \begin{itemize}
  \item{\it input:} $\alpha_s^{(n)}(\mu_0)$, $\mu_0$, $\mu$, number of loops
  \item{\it output:} $\alpha_s^{(m)}(\mu)$ 
  \item{\it uses:} {\tt AlphasExact[]}, {\tt AlL2AlH[]} and
    {\tt AlH2AlL[]}
  \item{\it comments:}
    The decoupling is performed automatically at the pole mass of the heavy
    quark where the values defined in {\tt Numdef} are taken.
    If $\mu$ is lower than $M_c$, $m=3$ is chosen, i.e. the strange quark is
    not decoupled.
  \item{\it example:} In order to compute
    $\alpha_s^{(6)}(500~\mbox{GeV})=0.952$ 
    with four-loop accuracy if
    $\alpha_s^{(5)}(M_Z)=0.118$ is given one has to use the command
    \verb|AsRunDec[asMz/.NumDef,Mz/.NumDef,500,4]|.
  \end{itemize}
\end{itemize}
\end{flushleft}

The conversion of the on-shell mass $M$ to the $\overline{\rm MS}$ mass, $m$,
can be computed with the help of the procedure {\tt mOS2mMS[]}:

\begin{flushleft}
\begin{itemize}
\item {\tt mOS2mMS}:
  \begin{itemize}
  \item{\it input:} $M$, $n_f$, number of loops
  \item{\it output:} $m^{(n_f)}(M)$ 
  \item{\it uses:} {\tt AsRunDec[]} and {\tt mOS2mMS[]} (from the appendix)
  \item{\it comments:} The relation is implemented up to order $\alpha_s^3$
    (three loops). $n_f$ is the number of active flavours.
    $\alpha_s^{(n_f)}(M)$ is evaluated at the scale $M$ where
    $\alpha_s^{(5)}(M_Z)$, as defined in {\tt Numdef}, 
    serves as a starting point. For the running and decoupling the
    procedure {\tt AsRunDec} is used.    
  \item{\it example:} In the case of the top quark the $\overline{\rm MS}$
    mass $m_t(M_t)=164.6$~GeV is obtained via
    \verb|mOS2mMS[175,6,3]| where $M_t=175$~GeV has been chosen.
  \end{itemize}
\end{itemize}
\end{flushleft}

\noindent
The inverted relation is implemented in

\begin{flushleft}
\begin{itemize}
\item {\tt mMS2mOS}:
  \begin{itemize}
  \item{\it input:} $\mu_m=m^{(n_f)}(\mu_m)$, $n_f$, 
    number of loops
  \item{\it output:} $M$ 
  \item{\it uses:} {\tt AsRunDec[]} and {\tt mMS2mOS[]} (from the appendix)
  \item{\it comments:} The relation is implemented up to order $\alpha_s^3$
    (three loops). $n_f$ is the number of active flavours.
    $\alpha_s^{(n_f)}(\mu_m)$ is evaluated at the scale $\mu_m$ where
    $\alpha_s^{(5)}(M_Z)$, as defined in {\tt Numdef}, 
    serves as a starting point. For the running and decoupling the
    procedure {\tt AsRunDec} is used.    
  \item{\it example:} In the case of the top quark the on-shell mass
    $M_t=174.7$~GeV is obtained via
    \verb|mMS2mOS[165,6,3]| where $m_t(M_t)=165$~GeV has been chosen.
  \end{itemize}
\end{itemize}
\end{flushleft}

In the above procedures all quarks lighter than the one under consideration
are assumed to be massless.
More specialized procedures providing more freedom in the choice of
parameters and the running presentations can be found in the Appendix.


\section*{Acknowledgments}

We would like to thank R.~Harlander for carefully reading the manuscript and
for many valuable comments. 
Also comments from G.~Rodrigo and
T.~van~Ritbergen are gratefully acknowledged.
This work was supported by DFG under Contract Ku 502/8-1
({\it DFG-Forschergruppe ``Quantenfeldtheorie, Computeralgebra und
  Monte-Carlo-Simulationen''}).


\section*{Appendix: Detailed presentation of the {\tt Mathematica} modules
  contained in {\tt RunDec}}

In the following we list the procedures contained in the program package
{\tt RunDec} and provide a brief description.
The order of the parameters specified in the field {\it input}
corresponds to the order required in the {\tt Mathematica} procedures.
The precision used for most of the numerical evaluations
is controlled with the variable
{\tt \$NumPrec}. For the procedures involving, e.g.,
numerical solutions of differential equations or recursive solutions of
equations the default precision of {\tt Mathematica} is kept which is
for all practical purposes more than enough.
Note that often the precision requested for with {\tt \$NumPrec}
can not be reached when the input data are only known to a few
digits.

{\tt AsRunDec[]} is not listed as it can already be found in
Section~\ref{sec:desc}. 

\subsection*{Procedures related to the strong coupling constant}


\begin{flushleft}
\begin{itemize}
\item {\tt LamExpl}:
  \begin{itemize}
  \item{\it input:} $\alpha_s^{(n_f)}(\mu)$, $\mu$, $n_f$, number of loops
  \item{\it output:} $\Lambda^{(n_f)}$
  \item{\it uses:} Eq.~(\ref{eq:lamexpl})
  \item{\it comments:} ---
  \item{\it example:} From the knowledge of
    $\alpha_s^{(5)}(M_Z)=0.118$ the computation of $\Lambda^{(5)}=0.2089$ to
    three-loop accuracy proceeds as follows:
    \verb|LamExpl[asMz/.NumDef,Mz/.NumDef,5,3]|.
  \end{itemize}

\item {\tt LamImpl}:
  \begin{itemize}
  \item{\it input:} $\alpha_s^{(n_f)}(\mu)$, $\mu$, $n_f$, number of loops
  \item{\it output:} $\Lambda^{(n_f)}$
  \item{\it uses:} Eq.~(\ref{eq:alp})
  \item{\it comments:} Solves Eq.~(\ref{eq:alp})
    numerically for $\Lambda^{(n_f)}$.
  \item{\it example:} If
    $\alpha_s^{(5)}(M_Z)=0.118$ is given
    the computation of $\Lambda^{(5)}=0.2083$ to
    three-loop accuracy proceeds as follows:
    \verb|LamImpl[asMz/.NumDef,Mz/.NumDef,5,3]|.
  \end{itemize}

\item {\tt AlphasLam}:
  \begin{itemize}
  \item{\it input:} $\Lambda^{(n_f)}$, $\mu$, $n_f$, number of loops
  \item{\it output:} $\alpha_s^{(n_f)}(\mu)$
  \item{\it uses:} Eq.~(\ref{eq:alp})
  \item{\it comments:} An explicit warning is printed on the screen
    if the ratio $\mu/\Lambda^{(n_f)}$ is too small.
  \item{\it example:} For $\Lambda^{(5)}=0.208$ and $M_b=4.7$~GeV the value of
    $\alpha_s^{(5)}(M_b)=0.2163$ is obtained to three-loop accuracy with 
    \verb|AlphasLam[0.208,4.7,5,3]|.
  \end{itemize}

\item {\tt AlphasExact}:
  \begin{itemize}
  \item{\it input:} $\alpha_s^{(n_f)}(\mu_0)$, $\mu_0$, $\mu$, $n_f$,
    number of loops
  \item{\it output:} $\alpha_s^{(n_f)}(\mu)$
  \item{\it uses:} Eq.~(\ref{eq:defbeta})
  \item{\it comments:} Solves the differential equation numerically using
    $\alpha_s(\mu_0)$ as initial condition.
    An explicit warning is printed on the screen
    if the ratio $\mu/\Lambda^{(n_f)}$ is too small
    where $\Lambda^{(n_f)}$ is obtained with the help of
    {\tt LamExpl[]}.
  \item{\it example:} $\alpha_s^{(5)}(M_b)=0.2167$ is computed from
    $\alpha_s^{(5)}(M_Z)$ through 
    \verb|AlphasExact[asMz/.NumDef,Mz/.NumDef,Mb/.NumDef,5,3]| where the
    three-loop formulae are used.
  \end{itemize}
\end{itemize}
\end{flushleft}


\subsection*{Procedures relating different mass definitions}

\begin{flushleft}
\begin{itemize}
\item {\tt mOS2mMS}:
  \begin{itemize}
  \item{\it input:} $M$, $\{\smM_q\}$, $\alpha_s^{(n_f)}(\mu)$, $\mu$, $n_f$,
    number of loops
  \item{\it output:} $m^{(n_f)}(\mu)$ 
  \item{\it uses:} Eq.~(\ref{eq:zmlog}) and Tab.~\ref{tab:nl}
  \item{\it comments:} The relation is implemented up to order $\alpha_s^3$
    (three loops). $\{\smM_q\}$. 
    is a set of light quark masses which can also be
    empty. For consistency reasons their values must correspond to the
    on-shell mass.

    Note that the name of the procedure is the same as the one introduced in
    Section~\ref{sec:desc}. The distinction is only in the number of the
    arguments.  
  \item{\it example:} The $\overline{\rm MS}$ mass corresponding to the
    on-shell top quark mass of 175~GeV is computed via 
    \verb|mOS2mMS[175,{},0.107,175,6,3]| where
    $\alpha_s^{(6)}(175~\mbox{GeV})=0.107$ has been chosen.
    The result reads $m_t(175~\mbox{GeV})=164.64$~GeV.
    Terms up to order $\alpha_s^3$ have been used and light quark mass effects
    have been neglected.
  \end{itemize}

\item {\tt mMS2mOS}:
  \begin{itemize}
  \item{\it input:} $m^{(n_f)}(\mu)$, $\{m_q\}$, $\alpha_s^{(n_f)}(\mu)$,
    $\mu$, $n_f$, number of loops
  \item{\it output:} $M$ 
  \item{\it uses:} Eq.~(\ref{eq:zminv}) for general $\mu$ 
    and Tab.~\ref{tab:nl}
  \item{\it comments:} The relation is implemented up to order $\alpha_s^3$
    (three loops). In this case the light quark masses are defined in the
    $\overline{\rm MS}$ scheme $\{m_q\}$.

    Note that the name of the procedure is the same as the one introduced in
    Section~\ref{sec:desc}. The distinction is only in the number of the
    arguments.  
  \item{\it example:} The on-shell mass corresponding to the
    $\overline{\rm MS}$ top quark mass $m_t(175~\mbox{GeV})=165$~GeV
    is computed via 
    \verb|mMS2mOS[165,{},0.107,175,6,3]| where
    $\alpha_s^{(6)}(175~\mbox{GeV})=0.107$ has been chosen.
    The result reads $M_t=175.35$~GeV.
    Terms up to order $\alpha_s^3$ have been used and light quark mass effects
    have been neglected.
  \end{itemize}

\item {\tt mOS2mMSrun}:
  \begin{itemize}
  \item{\it input:} $M$, $\{\smM_q\}$, $\alpha_s^{(n_f)}(\mu)$, $\mu$, $n_f$,
    number of loops
  \item{\it output:} $m^{(n_f)}(\mu)$ 
  \item{\it uses:} {\tt AlphasExact[]}, {\tt mOS2mSI[]} and {\tt mMS2mMS[]}
  \item{\it comments:} In a first step $\mu_m$ is computed and
    afterwards $m^{(n_f)}(\mu)$ is evaluated.
    The usage is identical to {\tt mOS2mMS[]}.
  \item{\it example:} (analog to {\tt mOS2mMS[]})
  \end{itemize}

\item {\tt mMS2mOSrun}:
  \begin{itemize}
  \item{\it input:} $m^{(n_f)}(\mu)$, $\{m_q\}$, $\alpha_s^{(n_f)}(\mu)$,
    $\mu$, $n_f$, number of loops
  \item{\it output:} $M$ 
  \item{\it uses:} {\tt AlphasLam[]}, {\tt LamImpl[]}, {\tt mMS2mMS[]} and
    {\tt mMS2mOS[]}
  \item{\it comments:} In a first step $\mu_m$ is computed. Then
    Eq.~(\ref{eq:zminv}) only has to be used for $\mu=\mu_m$.
    The usage is identical to {\tt mMS2mOS[]}.
  \item{\it example:} (analog to {\tt mMS2mOS[]})
  \end{itemize}

\item {\tt mOS2mMSit}:
  \begin{itemize}
  \item{\it input:} $M$, $\{m_q\}$, $\alpha_s^{(n_f)}(\mu)$, $\mu$, $n_f$,
    number of loops
  \item{\it output:} $m^{(n_f)}(\mu)$ 
  \item{\it uses:} Eq.~(\ref{eq:zminv}) for general $\mu$ 
    and Tab.~\ref{tab:nl}
  \item{\it comments:} For the computation
    Eq.~(\ref{eq:zminv}) is used in order to avoid the on-shell
    masses of the light quark masses $\{m_q\}$.
    The usage is identical to {\tt mOS2mMS[]}.
  \item{\it example:} (analog to {\tt mOS2mMS[]}).
  \end{itemize}

\item {\tt mOS2mSI}:
  \begin{itemize}
  \item{\it input:} $M$, $\{\smM_q\}$, $\alpha_s^{(n_f)}(M)$, $n_f$,
    number of loops
  \item{\it output:} $\mu_m=m^{(n_f)}(\mu_m)$ 
  \item{\it uses:} Eq.~(\ref{eq:zmzm}) and Tab.~\ref{tab:nl}
  \item{\it comments:} The scale invariant mass is computed
    from the on-shell mass.
  \item{\it example:} In the case of the bottom quark, the mass
    $\mu_b=3.97$~GeV is evaluated via
    \verb|mOS2mSI[Mb/.NumDef,{1.6},0.217,5,3]|
    where 
    $\alpha_s^{(5)}(M_b)=0.217$ has been chosen.
    In the mass relation terms up to order $\alpha_s^3$ have been used
    and quark mass effects arising from $M_c/M_b$ with $M_c=1.6$~GeV
    have been taken into account.
  \end{itemize}

\item {\tt mMS2mMS}:
  \begin{itemize}
  \item{\it input:} $m^{(n_f)}(\mu_0)$, $\alpha_s^{(n_f)}(\mu_0)$,
    $\alpha_s^{(n_f)}(\mu)$, $n_f$, number of loops
  \item{\it output:} $m^{(n_f)}(\mu)$ 
  \item{\it uses:} Eqs.~(\ref{eq:mrun}) and~(\ref{eq:mas})
  \item{\it comments:} ---
  \item{\it example:} From $m_b(M_b)=3.85$~GeV
    one finds $m_b(M_Z)=2.69$~GeV with the help of
    \verb|mMS2mMS[3.85,0.217,asMz/.NumDef,5,4])|
    where $\alpha_s^{(5)}(M_b)=0.217$ and $\alpha_s^{(5)}(M_Z)=0.217$
    has been used. For the running the four-loop expressions have been used.
  \end{itemize}

\item {\tt mMS2mSI}:
  \begin{itemize}
  \item{\it input:} $m^{(n_f)}(\mu)$, $\alpha_s^{(n_f)}(\mu)$,
    $\mu$, $n_f$, number of loops
  \item{\it output:} $\mu_m=m(\mu_m)$ 
  \item{\it uses:} Eqs.~(\ref{eq:mrun}) and~(\ref{eq:mas})
  \item{\it comments:} The scale invariant mass is computed
    from the $\overline{\rm MS}$ mass.
  \item{\it example:} $\mu_b=3.97$~GeV is computed from the input 
    $m_b(M_b)=3.85$~GeV, $M_b=4.7$~GeV and $\alpha_s^{(5)}(M_b)=0.217$
    via the command
    \verb|mMS2mSI[3.85,0.217,4.7,5,4]|.
    For the running the four-loop expressions have been used.
  \end{itemize}

\item {\tt mMS2mRI}:
  \begin{itemize}
  \item{\it input:} $m^{(n_f)}(\mu)$, $\alpha_s^{(n_f)}(\mu)$,
    $n_f$, number of loops
  \item{\it output:} $m^{RI}$ 
  \item{\it uses:} inverted equation of~(\ref{eq:RI2MS})
  \item{\it comments:} The relation is implemented up to order $\alpha_s^3$
    (three loops).
  \item{\it example:} The regularization invariant mass, $m^{RI}_b(M_Z)$,
    corresponding
    to the $\overline{\rm MS}$ bottom quark mass $m_b(M_Z)=2.695$~GeV
    is computed via 
    \verb|mMS2mRI[2.695,asMz/.NumDef,5,3]| where
    $\alpha_s^{(5)}(M_Z)=0.118$ has been chosen.
    The result reads $m^{RI}_b(M_Z)=2.872$~GeV where
    terms up to order $\alpha_s^3$ have been used.
  \end{itemize}

\item {\tt mRI2mMS}:
  \begin{itemize}
  \item{\it input:} $m^{RI}(\mu)$, $\alpha_s^{(n_f)}(\mu)$,
    $n_f$, number of loops
  \item{\it output:} $m^{(n_f)}(\mu)$ 
  \item{\it uses:} Eq.~(\ref{eq:RI2MS})
  \item{\it comments:} The relation is implemented up to order $\alpha_s^3$
    (three loops).
  \item{\it example:} The $\overline{\rm MS}$ mass corresponding to the
    regularization invariant top quark mass of 
    $m_t^{RI}=175$~GeV is computed via 
    \verb|mRI2mMS[175,0.107,175,6,3]| where
    $\alpha_s^{(6)}(175~\mbox{GeV})=0.107$ has been chosen.
    The result reads $m_t(175~\mbox{GeV})=165.6$~GeV where
    terms up to order $\alpha_s^3$ have been used.
  \end{itemize}

\item {\tt mMS2mRGI}:
  \begin{itemize}
  \item{\it input:} $m^{(n_f)}(\mu)$, $\alpha_s^{(n_f)}(\mu)$,
    $n_f$, number of loops
  \item{\it output:} $\hat{m}$ 
  \item{\it uses:} Eq~(\ref{eq:mhat})
  \item{\it comments:} ---
  \item{\it example:} The renormalization group invariant bottom quark mass
    corresponding to the $\overline{\rm MS}$ mass $m_b^{(5)}(M_Z)=2.69$~GeV
    is computed via \verb|mMS2mRGI[2.69,asMz/.NumDef,5,4]| where
    $\alpha_s^{(5)}(M_Z)=0.118$ has been chosen.
    The result reads $m_b^{RGI}=14.25$~GeV assuming four-loop accuracy.
  \end{itemize}

\item {\tt mRGI2mMS}:
  \begin{itemize}
  \item{\it input:} $\hat{m}$, $\alpha_s^{(n_f)}(\mu)$,
    $n_f$, number of loops
  \item{\it output:} $m^{(n_f)}(\mu)$ 
  \item{\it uses:} Eq~(\ref{eq:mhat})
  \item{\it comments:} ---
  \item{\it example:} The $\overline{\rm MS}$ mass corresponding to the
    renormalization group invariant bottom quark mass of 14.25~GeV
    is computed via 
    \verb|mRGI2mMS[14.25,asMz/.NumDef,5,4]| where
    $\alpha_s^{(5)}(M_Z)=0.118$ has been chosen.
    The result reads $m_b(M_Z)=2.69$~GeV assuming four-loop accuracy.
  \end{itemize}

\end{itemize}
\end{flushleft}


\subsection*{Decoupling of the strong coupling and the masses}

At this point we once 
again want to stress, that the argument specifying the number
of loops refers to the accompanied running, i.e. if
``2'' is chosen the decoupling relation is used to one-loop order. 
Furthermore, for the argument ruling the number of active flavours the number
of light quarks, $n_l=n_f-1$, is chosen.

\begin{flushleft}
\begin{itemize}
\item {\tt DecAsUpOS}:
  \begin{itemize}
  \item{\it input:} $\alpha_s^{(n_l)}(\mu_{th})$, $M_{th}$, $\mu_{th}$,
    $n_l$, number of loops
  \item{\it output:} $\alpha_s^{(n_l+1)}(\mu_{th})$
  \item{\it uses:} Eq.~(\ref{eq:invzetagOS})
  \item{\it comments:} For the heavy mass the on-shell definition is used.
  \item{\it example:} The computation of $\alpha_s^{(6)}(M_Z)=0.1169$
    from the knowledge of $\alpha_s^{(5)}(M_Z)=0.118$ proceeds via
    \verb|DecAsUpOS[asMz/.NumDef,175,Mz/.NumDef,5,4]| where $M_t=175$~GeV
    has been chosen and terms of order $\alpha_s^3$ (indicated by the ``4''
    in the last argument) have been included.    
  \end{itemize}

\item {\tt DecAsDownOS}:
  \begin{itemize}
  \item{\it input:} $\alpha_s^{(n_l+1)}(\mu_{th})$, $M_{th}$, $\mu_{th}$,
    $n_l$, number of loops
  \item{\it output:} $\alpha_s^{(n_l)}(\mu_{th})$
  \item{\it uses:} Eq.~(\ref{eq:zetagOS})
  \item{\it comments:} For the heavy mass the on-shell definition is used.
  \item{\it example:} The computation of
    $\alpha_s^{(5)}(200~\mbox{GeV})=0.1047$
    from the knowledge of $\alpha_s^{(6)}(200~\mbox{GeV})=0.105$ proceeds via
    \verb|DecAsDownOS[0.105,175,200,6,4]| where $M_t=175$~GeV
    has been chosen and terms of order $\alpha_s^3$ (indicated by the ``4''
    in the last argument) have been included.    
  \end{itemize}

\item {\tt DecAsUpMS}:
  \begin{itemize}
  \item{\it input:} $\alpha_s^{(n_l)}(\mu_{th})$, $m_{th}(\mu_{th})$, 
    $\mu_{th}$, $n_l$, number of loops
  \item{\it output:} $\alpha_s^{(n_l+1)}(\mu_{th})$
  \item{\it uses:} Eq.~(\ref{eq:invzetagMS})
  \item{\it comments:} The heavy mass is evaluated in the $\overline{\rm MS}$
    scheme at the scale $\mu_{th}$.
  \item{\it example:} The computation of $\alpha_s^{(6)}(M_Z)=0.1170$
    from the knowledge of $\alpha_s^{(5)}(M_Z)=0.118$ proceeds via
    \verb|DecAsUpMS[asMz/.NumDef,165,Mz/.NumDef,5,4]| where $m_t(M_Z)=165$~GeV
    has been chosen and terms of order $\alpha_s^3$ (indicated by the ``4''
    in the last argument) have been included.    
  \end{itemize}

\item {\tt DecAsDownMS}:
  \begin{itemize}
  \item{\it input:} $\alpha_s^{(n_l+1)}(\mu_{th})$, $m_{th}(\mu_{th})$, 
    $\mu_{th}$, $n_l$, number of loops
  \item{\it output:} $\alpha_s^{(n_l)}(\mu_{th})$
  \item{\it uses:} Eq.~(\ref{eq:zetagMS})
  \item{\it comments:} The heavy mass is evaluated in the $\overline{\rm MS}$
    scheme at the scale $\mu_{th}$.
  \item{\it example:} The computation of
    $\alpha_s^{(5)}(200~\mbox{GeV})=0.1048$
    from the knowledge of $\alpha_s^{(6)}(200~\mbox{GeV})=0.105$ proceeds via
    \verb|DecAsDownMS[0.105,165,200,6,4]| where $m_t(M_Z)=165$~GeV
    has been chosen and terms of order $\alpha_s^3$ (indicated by the ``4''
    in the last argument) have been included.    
  \end{itemize}

\item {\tt DecAsUpSI}:
  \begin{itemize}
  \item{\it input:} $\alpha_s^{(n_l)}(\mu_{th})$, $\mu_{m_{th}}$, 
    $\mu_{th}$, $n_l$, number of loops
  \item{\it output:} $\alpha_s^{(n_l+1)}(\mu_{th})$
  \item{\it uses:} Eq.~(\ref{eq:invzetagSI})
  \item{\it comments:} Here the scale invariant mass
    $\mu_{m_{th}}$ is chosen for heavy mass.
  \item{\it example:} (analog to {\tt DecAsUpMS[]})
  \end{itemize}

\item {\tt DecAsDownSI}:
  \begin{itemize}
  \item{\it input:} $\alpha_s^{(n_l+1)}(\mu_{th})$, $\mu_{m_{th}}$, 
    $\mu_{th}$, $n_l$, number of loops
  \item{\it output:} $\alpha_s^{(n_l)}(\mu_{th})$
  \item{\it uses:} Eq.~(\ref{eq:zetagSI})
  \item{\it comments:} Here the scale invariant mass
    $\mu_{m_{th}}$ is chosen for heavy mass.
  \item{\it example:} (analog to {\tt DecAsDownMS[]})
  \end{itemize}

\item {\tt DecMqUpOS}:
  \begin{itemize}
  \item{\it input:} $m_q^{(n_l)}(\mu_{th})$, $\alpha_s^{(n_l)}(\mu_{th})$,
    $M_{th}$, $\mu_{th}$, $n_l$, number of loops
  \item{\it output:} $m_q^{(n_l+1)}(\mu_{th})$
  \item{\it uses:} Eq.~(\ref{eq:invzetamOS})
  \item{\it comments:} For the heavy mass the on-shell definition is used.
  \item{\it example:} The computation of $m_b^{(6)}(M_Z)=2.697$~GeV from
    $m_b^{(5)}(M_Z)=2.7$~GeV with order $\alpha_s^3$ accuracy
    is performed via
    \verb|DecMqUpOS[2.7,asMz/.NumDef,175,Mz/.NumDef,5,4]|.
    Here, $\alpha_s^{(5)}(M_Z)=0.118$ and $M_t=175$~GeV have been used.
  \end{itemize}

\item {\tt DecMqDownOS}:
  \begin{itemize}
  \item{\it input:} $m_q^{(n_l+1)}(\mu_{th})$, $\alpha_s^{(n_l+1)}(\mu_{th})$,
    $M_{th}$, $\mu_{th}$, $n_l$, number of loops
  \item{\it output:} $m_q^{(n_l)}(\mu_{th})$
  \item{\it uses:} Eq.~(\ref{eq:zetamOS})
  \item{\it comments:} For the heavy mass the on-shell definition is used.
  \item{\it example:} The computation of $m_c^{(4)}(M_Z)=0.583$~GeV from
    $m_c^{(5)}(M_Z)=0.58$~GeV with order $\alpha_s^3$ accuracy
    is performed via
    \verb|DecMqDownOS[0.58,asMz/.NumDef,4.7,Mz/.NumDef,5,4]|.
    Here, $\alpha_s^{(5)}(M_Z)=0.118$ and $M_b=4.7$~GeV have been used.
  \end{itemize}

\item {\tt DecMqUpMS}:
  \begin{itemize}
  \item{\it input:} $m_q^{(n_l)}(\mu_{th})$, $\alpha_s^{(n_l)}(\mu_{th})$,
    $m_{th}(\mu_{th})$, $\mu_{th}$, $n_l$, number of loops
  \item{\it output:} $m_q^{(n_l+1)}(\mu_{th})$
  \item{\it uses:} Eq.~(\ref{eq:invzetamMS})
  \item{\it comments:} The heavy mass is evaluated in the $\overline{\rm MS}$
    scheme at the scale $\mu_{th}$.
  \item{\it example:} (analog to {\tt DecMqUpOS[]})
  \end{itemize}

\item {\tt DecMqDownMS}:
  \begin{itemize}
  \item{\it input:} $m_q^{(n_l+1)}(\mu_{th})$, $\alpha_s^{(n_l+1)}(\mu_{th})$,
    $m_{th}(\mu_{th})$, $\mu_{th}$, $n_l$, number of loops
  \item{\it output:} $m_q^{(n_l)}(\mu_{th})$
  \item{\it uses:} Eq.~(\ref{eq:zetamMS})
  \item{\it comments:} The heavy mass is evaluated in the $\overline{\rm MS}$
    scheme at the scale $\mu_{th}$.
  \item{\it example:} (analog to {\tt DecMqDownOS[]})
  \end{itemize}

\item {\tt DecMqUpSI}:
  \begin{itemize}
  \item{\it input:} $m_q^{(n_l)}(\mu_{th})$, $\alpha_s^{(n_l)}(\mu_{th})$,
    $\mu_{m_th}$, $\mu_{th}$, $n_l$, number of loops
  \item{\it output:} $m_q^{(n_l+1)}(\mu_{th})$
  \item{\it uses:} Eq.~(\ref{eq:invzetamSI})
  \item{\it comments:} Here the scale invariant mass
    $\mu_{m_{th}}$ is chosen for heavy mass.
  \item{\it example:} (analog to {\tt DecMqUpOS[]})
  \end{itemize}

\item {\tt DecMqDownSI}:
  \begin{itemize}
  \item{\it input:} $m_q^{(n_l+1)}(\mu_{th})$, $\alpha_s^{(n_l+1)}(\mu_{th})$,
    $\mu_{m_th}$, $\mu_{th}$, $n_l$, number of loops
  \item{\it output:} $m_q^{(n_l)}(\mu_{th})$
  \item{\it uses:} Eq.~(\ref{eq:zetamSI})
  \item{\it comments:} Here the scale invariant mass
    $\mu_{m_{th}}$ is chosen for heavy mass.
  \item{\it example:} (analog to {\tt DecMqUpMS[]})
  \end{itemize}

\item {\tt DecLambdaUp}:
  \begin{itemize}
  \item{\it input:} $\Lambda^{(n_l)}$, $\mu_{m_th}$, $n_l$, number of loops
  \item{\it output:} $\Lambda^{(n_l+1)}$
  \item{\it uses:} Eq.~(\ref{eq:declam}) 
  \item{\it comments:} For the heavy mass the scale
    invariant mass $\mu_{m_{th}}$ is used.
  \item{\it example:} From $\Lambda^{(4)}=0.2876$ one can compute
    $\Lambda^{(5)}=0.208$ with the help of
    \verb|DecLambdaUp[0.287,3.97,4,4]|
    where $\mu_b=3.97$ and four-loop accuracy has been chosen.
  \end{itemize}

\item {\tt DecLambdaDown}:
  \begin{itemize}
  \item{\it input:} $\Lambda^{(n_l+1)}$, $\mu_{m_th}$, $n_l$, number of loops
  \item{\it output:} $\Lambda^{(n_l)}$
  \item{\it uses:} Eq.~(\ref{eq:declam2}) 
  \item{\it comments:} For the heavy mass the scale
    invariant mass $\mu_{m_{th}}$ is used.
  \item{\it example:} From $\Lambda^{(5)}=0.208$ one can compute
    $\Lambda^{(4)}=0.208$ with the help of
    \verb|DecLambdaDown[0.208,3.97,4,4]|
    where $\mu_b=3.97$ and four-loop accuracy has been chosen.
  \end{itemize}
\end{itemize}
\end{flushleft}


\subsection*{Miscellaneous procedures}

The following modules provide some simple examples which mostly combine the
modules described above. ``L'' stands for low and ``H'' for high. 
The condition $l<h$ is  assumed in  all four procedures.  

\begin{flushleft}
\begin{itemize}
\item {\tt AlL2AlH}:
  \begin{itemize}
  \item{\it input:} $\alpha_s^{(l)}(\mu_1)$, $\mu_1$,
    $\{\{n_{f_1},M_{th_1},\mu_{th_1}\},\{n_{f_2},M_{th_2},\mu_{th_2}\},
    \ldots\}$, $\mu_2$, number of loops
  \item{\it output:} $\alpha_s^{(h)}(\mu_2)$
  \item{\it uses:} {\tt AlphasExact[]} and {\tt DecAsUpOS[]}
  \item{\it comments:}
    The set in the third argument
    may contain several triples  indicating the
    number of flavours, the heavy (on-shell) 
    quark mass and the scale at which the
    decoupling is performed.
  \item{\it examples:} 
    1. For the computation of $\alpha_s^{(6)}(500~\mbox{GeV})=0.0952$
    from $\alpha_s^{(4)}(M_c=1.6~\mbox{GeV})=0.338$
    to ${\cal O}(\alpha_s^3)$ accuracy the
    input would look as follows:
    \verb|AlL2AlH[0.338,1.6,{{5,4.7,5},{6,175,200}},500,4]|
    Here, the matching is performed at 5~GeV and 200~GeV, respectively.

    2. Fig.~\ref{fig:asMZasMtau} can be reproduced with the help of the
    following input
    \verb|AlL2AlH[0.36,1.777,{{5,4.7,mu5}},91.187,l]|
    where $l=1,2,3,4$ corresponds to the number of loops and \verb|mu5/4.7|
    is the scale on the abscissa.
  \end{itemize}

\item {\tt AlH2AlL}:
  \begin{itemize}
  \item{\it input:} $\alpha_s^{(h)}(\mu_1)$, $\mu_1$,
    $\{\{n_{f_1},M_{th_1},\mu_{th_1}\},\{n_{f_2},M_{th_2},\mu_{th_2}\},
    \ldots\}$, $\mu_2$, number of loops
  \item{\it output:} $\alpha_s^{(l)}(\mu_2)$
  \item{\it uses:} {\tt AlphasExact[]} and {\tt DecAsDownOS[]}
  \item{\it comments:} The set in the third argument
    may contain several triples  indicating the
    number of flavours, the heavy (on-shell) 
    quark mass and the scale at which the
    decoupling is performed.
  \item{\it example:} Consider the inverse order of the first example of
    the previous procedure. The input 
    \verb|AlH2AlL[0.0952,500,{{6,175,200},{5,4.7,5}},1.6,4]|,
    indeed leads to $\alpha_s^{(4)}(1.6~\mbox{GeV})=0.338$. 
  \end{itemize}

\item {\tt mL2mH}:
  \begin{itemize}
  \item{\it input:} $m_q^{(l)}(\mu_1)$, $\alpha_s^{(l)}(\mu_1)$, $\mu_1$,
    $\{\{n_{f_1},M_{th_1},\mu_{th_1}\},\{n_{f_2},M_{th_2},\mu_{th_2}\},
    \ldots\}$, $\mu_2$, number of loops
  \item{\it output:} $m_q^{(h)}(\mu_2)$
  \item{\it uses:} {\tt AlphasExact[]}, {\tt mMS2mMS[]} {\tt DecMqUpOS[]}
    and {\tt DecAsUpOS[]}
  \item{\it comments:} The set in the fourth argument
    may contain several triples  indicating the
    number of flavours, the heavy (on-shell) 
    quark mass and the scale at which the
    decoupling is performed.
  \item{\it example:} Using $\alpha_s^{(4)}(1.2~\mbox{GeV})=0.403$ and
    $m_c^{(4)}(1.2~\mbox{GeV})=1.2$~GeV
    one finds $m_c^{(5)}(M_Z)=0.580$~GeV with the help of
    \verb|mL2mH[1.2,0.403,1.2,{{5,4.7,5.0}},Mz/.NumDef,4]|.
    The decoupling of $M_b=4.7$~GeV is performed at $5.0$~GeV.
    In this way the results of Fig.~\ref{fig:mcMZ} can be reproduced.
  \end{itemize}

\item {\tt mH2mL}:
  \begin{itemize}
  \item{\it input:} $m_q^{(h)}(\mu_1)$, $\alpha_s^{(h)}(\mu_1)$, $\mu_1$,
    $\{\{n_{f_1},M_{th_1},\mu_{th_1}\},\{n_{f_2},M_{th_2},\mu_{th_2}\},
    \ldots\}$, $\mu_2$, number of loops
  \item{\it output:} $m_q^{(l)}(\mu_2)$
  \item{\it uses:} {\tt AlphasExact[]}, {\tt mMS2mMS[]} {\tt DecMqDownOS[]}
    and {\tt DecAsDownOS[]}
  \item{\it comments:} The set in the fourth argument
    may contain several triples  indicating the
    number of flavours, the heavy (on-shell) 
    quark mass and the scale at which the
    decoupling is performed.
  \item{\it example:}
    Using $\alpha_s^{(5)}(M_Z)=0.118$ and
    $m_c^{(5)}(M_Z)=0.580$~GeV
    one finds $m_c^{(4)}(1.2~\mbox{GeV})=1.20$~GeV with the help of
    \verb|mH2mL[0.580,asMz/.NumDef,Mz/.NumDef,{{5,4.7,5.0}},1.2,4]|.
    The decoupling of $M_b=4.7$~GeV is performed at $5.0$~GeV.    
  \end{itemize}

\end{itemize}
\end{flushleft}


\end{document}